\documentclass[referee,sn-nature]{sn-jnl}

\usepackage{graphicx}%
\usepackage{multirow}%
\usepackage{amsmath,amssymb,amsfonts}%
\usepackage{amsthm}%
\usepackage{mathrsfs}%
\usepackage[title]{appendix}%
\usepackage{xcolor}%
\usepackage{textcomp}%
\usepackage{manyfoot}%
\usepackage{booktabs}%
\usepackage{algorithm}%
\usepackage{algorithmicx}%
\usepackage{algpseudocode}%
\usepackage{listings}%
\usepackage{fullpage}
\usepackage{pdfpages}
\usepackage{float}

\usepackage{soul}
\DeclareUnicodeCharacter{0301}{\'{e}}
\begin{document}
\title[Article Title]{Emergent electronic landscapes in a novel valence-ordered nickelate with tri-component nickel coordination}
\author[1,2,9]{\fnm{Aravind} \sur{Raji}}
\author[3,4,9]{\fnm{Zhengang} \sur{Dong}}
\author[2]{\fnm{Victor} \sur{Por\'{e}e}}
\author[5]{\fnm{Alaska} \sur{Subedi}}
\author[1]{\fnm{Xiaoyan} \sur{Li}}
\author[6,7]{\fnm{Bernat} \sur{Mundet}}
\author[6]{\fnm{Lucia} \sur{Varbaro}}
\author[6]{\fnm{Claribel} \sur{Dom\'{i}nguez}}
\author[6]{\fnm{Marios} \sur{Hadjimichael}}
\author[3,4]{\fnm{Bohan} \sur{Feng}}
\author[2]{\fnm{Alessandro} \sur{Nicolaou}}
\author[2,8]{\fnm{Jean-Pascal} \sur{Rueff}}
\author*[3,4]{\fnm{Danfeng} \sur{Li}}\email{danfeng.li@cityu.edu.hk}
\author*[1]{\fnm{Alexandre} \sur{Gloter}}\email{alexandre.gloter@universite-paris-saclay.fr}

\affil[1]{\orgname{Laboratoire de Physique des Solides, CNRS, Universit\'{e} Paris-Saclay}, \orgaddress{ \city{Orsay}, \postcode{91400}, \country{France}}}
\affil[2]{\orgname{Synchrotron SOLEIL}, \orgaddress{\street{L’Orme des Merisiers, BP 48 St Aubin}, \city{Gif sur Yvette}, \postcode{91192}, \country{France}}}
\affil[3]{\orgdiv{Department of Physics}, \orgname{City University of Hong Kong}, \orgaddress{\city{Kowloon}, \country{Hong Kong}}}

\affil[4]{\orgdiv{City University of Hong Kong Shenzhen Research Institute}, \orgname{Shenzhen}, \orgaddress{\city{Guangdong}, \postcode{518057}, \country{China}}}

\affil[5]{\orgname{CPHT, Ecole Polytechnique}, \orgaddress{\city{Palaiseau cedex}, \postcode{91128}, \country{France}}}

\affil[6]{\orgname{Department of Quantum Matter Physics, University of Geneva}, \orgaddress{\city{Geneva}, \country{Switzerland}}}

\affil[7]{\orgname{Electron Spectrometry and Microscopy Laboratory (LSME), Institute of Physics (IPHYS), Ecole Polytechnique Fédérale de Lausanne (EPFL)}, \orgaddress{\city{Lausanne}, \country{Switzerland}}}

\affil[8]{\orgname{LCPMR, Sorbonne Université, CNRS}, \orgaddress{\city{Paris}, \postcode{75005}, \country{France}}}

\affil[9]{These authors contributed equally to this work}


\abstract{The metal-hydride-based ‘topochemical reduction’ process has produced novel thermodynamically unstable phases across various transition metal oxide series with unusual crystal structures and non-trivial ground states. Here, by such an oxygen (de-) intercalation method we synthesis a novel samarium nickelate with ordered nickel valences associated with tri-component coordination configurations. This structure, with a formula of Sm$_{9}$Ni$_{9}$O$_{22}$ as revealed by four-dimensional scanning transmission electron microscopy, emerges from the intricate planes of \{303\}$_{\text{pc}}$ ordered apical oxygen vacancies. X-ray spectroscopy measurements and ab-initio calculations show the coexistence of square-planar, pyramidal and octahedral Ni sites with mono-, bi- and tri-valences. It leads to an intense orbital polarization, charge-ordering, and a ground state with a strong electron localization marked by the disappearance of ligand-hole configuration at low-temperature. This new nickelate compound provides another example of previously inaccessible materials enabled by topotactic transformations and presents a unique platform where mixed Ni valence can give rise to exotic phenomena.}



\maketitle
\section*{Introduction}\label{sec1}
The on-demand design of transition-metal oxides (TMO) with emerging properties is imparted by the multivalent nature of the transition-metal ions and the accessible complexity of lattice structures \cite{dagotto2005complexity,ahn2021designing}. Varying the structure and constituent elements in TMO allows access to a range of competing ground states and exotic electronic landscapes, in which the interplay of the transition-metal orbitals and the connecting oxygen network often plays an essential role \cite{cheong2007exciting}. One phenomenal example is the cuprate family, where different coordinational Cu-O frameworks lead to distinctive $T_\textrm{c}$ ($T_\textrm{c}$, superconducting transition temperature) values \cite{keimer2015quantum}. In addition, the fast developments in topotactic soft-chemistry approaches have enabled the synthesis of unconventional TMO with unusual valence states and/or oxygen connectivity \cite{hayward2014topochemical,sanjaya2011topochemical,wu2023electrostatic}. One of the recent seminal examples is the synthesis of a family of superconducting low-valence nickelates \cite{li2019superconductivity,zeng2020phase,pan2022superconductivity,wei2023superconducting} by a topotactic reduction process, which displays intriguing analogies and distinctions to the high-$T_\textrm{c}$ cuprates and is the focus of the current extensive research. 

Nickel oxides or nickelates form a large family of structures with diverse crystalline configurations and electronic properties.
Perovskite rare-earth nickelates RNiO$_{3}$ have a rich phase diagram generated by changing the rare-earth cation. They exhibit metal-to-insulator, magnetic and structural transitions \cite{kumah2014tuning, zhang2018perovskite, mercy2017structurally}, and upon stoichiometry alteration such as oxygen deintercalation or protonization can become reconfigurable neuromorphic devices \cite{zhang2022reconfigurable}. Upon a strong topotactic oxygen reduction, RNiO$_{2}$ infinite-layer nickelates can be obtained, which have recently been discovered as superconducting after hole-doping  \cite{li2019superconductivity}.
While the origin of most of these properties is not yet fully understood, the interplay of spin, charge and orbital degrees of freedom is known to drive these exotic properties, along with the associated crystalline configurations.

In the case of infinite-layer nickelate superconductors, a structurally-perfect infinite-layer thin film is essential to achieve superconductivity \cite{lee2020aspects}. 
The topotactic reduction method for infinite-layer nickelate synthesis involves metal hydride, e.g. CaH$_{2}$, to remove the apical oxygens from a perovskite nickelate. This often results in a mixture of secondary phases with partial removal of apical oxygen \cite{wu2023topotactically, osada2023improvement}. This is due to the partial instability attributed to the infinite-layer structure that does not host Ni$^{2+}$ (3d$^{8}$), but a Ni$^{1+}$ in a 3d$^{9}$ configuration \cite{subedi2023possible}.
Along with superconductivity, the observed charge ordering (CO) \cite{krieger2022charge, ren2023symmetry, tam2022charge, rossi2022broken} in these systems now appears to be connected to a structural reordering due to oxygen intercalation in the apical sites \cite{raji2023charge,parzyck2023absence}. Nevertheless, the structure of the charge-ordered systems cannot be directly related to known phases, such as brownmillerite \cite{wu2023topotactically} or octahedra and square planar bearing R$_{3}$Ni$_{3}$O$_{7}$ phase where Ni$^{2+}$ is absent \cite{moriga1994reduction}. The complex transport properties and notably the unusual dramatic resistivity increase upon oxygen deintercalation of the perovskite nickelates RNiO$_{3}$ is also intriguing \cite{kotiuga2019carrier}. 

This concomitance indicates the possible existence of an unknown stable intermediate phase with multivalent, strong electron localisation and charge ordered ground state between the perovskite ABO$_{3}$ and the infinite-layer ABO$_{2}$.  

Here, we demonstrate that the topochemical reduction of a perovskite SmNiO$_{3}$ thin film grown on an orthorhombic NdGaO$_{3}$ substrate stabilizes to such an intermediate phase whose stoichiometry is derived as Sm$_{9}$Ni$_{9}$O$_{22}$. This ubiquitous phase existing between the much pronounced perovskite ABO$_{3}$ and the infinite-layer ABO$_{2}$ nickelates hosts intriguing properties associated with multi-valent and multi-component coordinated Ni sites. Intriguing properties of this phase include a strong-orbital polarization, temperature dependent carrier localization, and charge ordering. The observed structural and electronic properties from microscopy and spectroscopy are validated by ab-initio simulations, contributing to a coherent analysis and uncovering some regions of its possibly rich phase-diagram.       

\section*{Results}\label{sec2}
\subsection*{A new phase from the \{303\}$_{\text{pc}}$ apical oxygen vacancy ordering}
\begin{figure*}[h!]
\centering
\includegraphics[width=\textwidth]{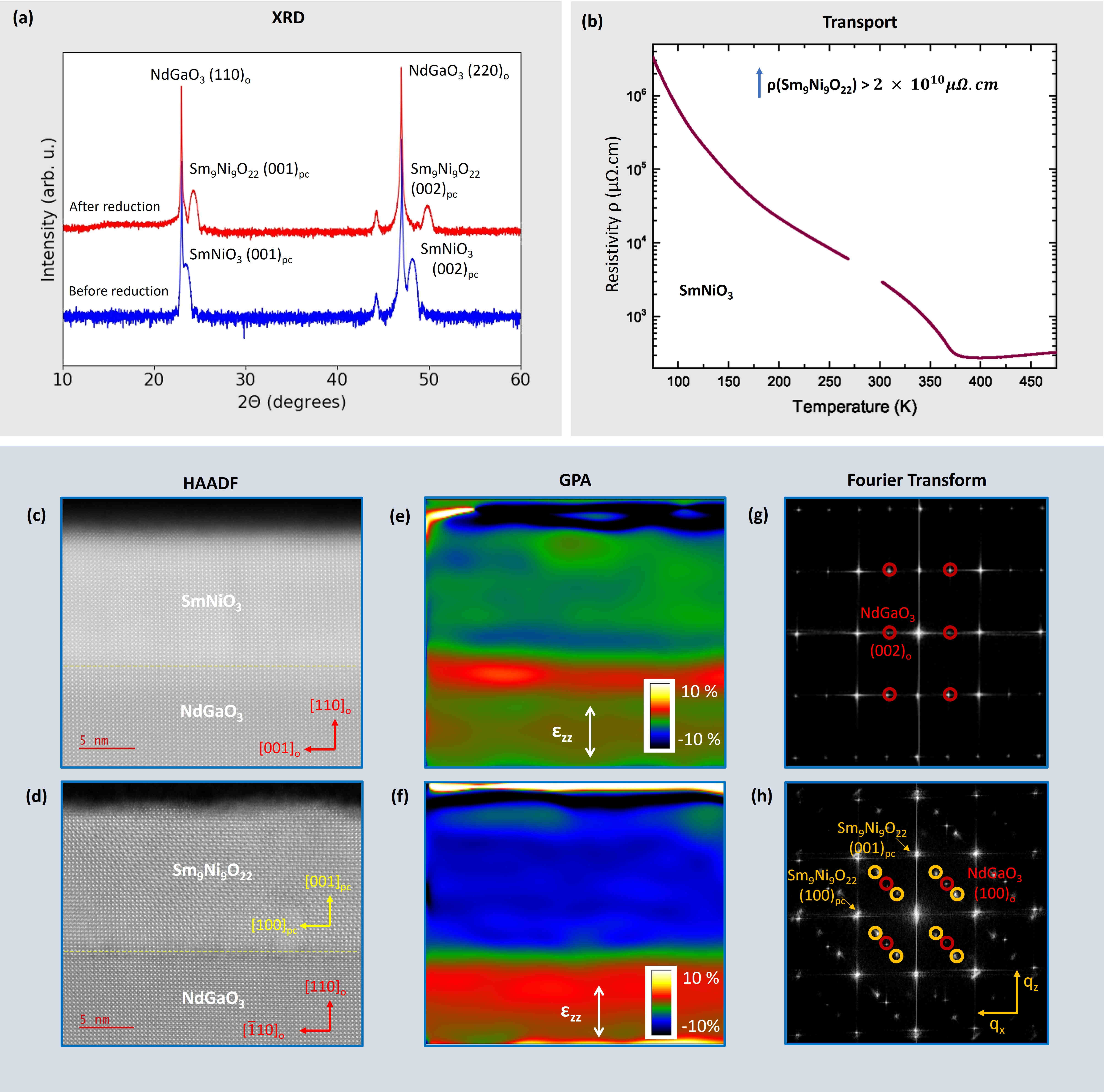}
\caption{Structural and transport comparison of SmNiO$_{3}$ with Sm$_{9}$Ni$_{9}$O$_{22}$ by XRD, HAADF-STEM imaging, GPA, and FT. (a) XRD comparison of both. The 001 and 002 peaks for SmNiO$_{3}$ occur at lower 2$\theta$ values compared to the 001 and 002 peaks of Sm$_{9}$Ni$_{9}$O$_{22}$, indicating an out-of-plane compression of the pseudocubic (pc) unit cell upon reduction. (b) Transport measurement of parent SmNiO$_{3}$ in increasing and decreasing temperatures. The resistivity curve of Sm$_{9}$Ni$_{9}$O$_{22}$ is not included in the plot since it was significantly higher, and is beyond the detection range of the equipment ($> 2 \times 10^{10}$ $\mu\Omega$.cm). (c) HAADF-STEM image of SmNiO$_{3}$ thin film. (d) HAADF-STEM image of Sm$_{9}$Ni$_{9}$O$_{22}$ thin film, showing in good contrast the stripes coinciding with the (303)$_{\text{pc}}$ direction. (e \& f) Maps of the o-o-p strain by GPA in (e) SmNiO$_{3}$ and (f) Sm$_{9}$Ni$_{9}$O$_{22}$. Both samples show uniform o-o-p compression. The Sm$_{9}$Ni$_{9}$O$_{22}$ exhibits a more negative $\epsilon_{zz}$ compared to SmNiO$_{3}$, corresponding to a lower o-o-p parameter. (g \& h) The corresponding FT of the HAADF image of (g) SmNiO$_{3}$, (h) Sm$_{9}$Ni$_{9}$O$_{22}$. The NGO substrate is indexed with an orthorhombic unit cell and the super-structure spots associated to the tilt pattern of the NGO are encircled in red. The super-structure spots associated to the (303)$_{\text{pc}}$ stripes in Sm$_{9}$Ni$_{9}$O$_{22}$ are encircled in yellow and are indexed within a pseudo-cubic unit cell.}
\label{fig:Figure 1}
\end{figure*}

We begin with a macroscopic comparison between the parent perovskite SmNiO$_{3}$ and the reduced Sm$_{9}$Ni$_{9}$O$_{22}$ thin-films, using x-ray diffraction (XRD) and transport measurements. As shown in Fig. \ref{fig:Figure 1}a, we find that the reduction of SmNiO$_{3}$ results in a scaling down of the out-of-plane (o-o-p) lattice parameter from 3.78 \AA\ to 3.63 \AA, indicative of a structural transition to the reduced Sm$_{9}$Ni$_{9}$O$_{22}$ phase. The thin-film quality is maintained, as evidenced by the intense Bragg peaks and finite-size oscillations for both phases. Additionally, in the transport measurements shown in Fig.\ref{fig:Figure 1}b, SmNiO$_{3}$ evidences a semiconducting behavior with an increase in conductivity of 4 orders of magnitude between 75K and room-temperature, and a metal-to-insulator transition around 375K. This is in agreement with previous reports on SmNiO$_{3}$ thin-films \cite{torriss2017metal}. On the other hand, the resistivity of Sm$_{9}$Ni$_{9}$O$_{22}$ was above the measurement range  $> 2 \times 10^{10}$ $\mu$$\Omega$.cm making it highly insulating with respect to its parent phase. 

The microscopic characterization begins with a structural comparison between the parent perovskite SmNiO$_{3}$ and the reduced Sm$_{9}$Ni$_{9}$O$_{22}$ thin-films, using high-angle annular dark-field (HAADF)-STEM imaging, geometrical phase analysis (GPA) \cite{hytch1998quantitative}, and Fourier transform (FT) analysis. As shown in Fig.\ref{fig:Figure 1}c \& d, the HAADF-STEM image shows homogeneous thin films for both SmNiO$_{3}$ and Sm$_{9}$Ni$_{9}$O$_{22}$, with a coherent growth obtained on the (110)$_{\text{o}}$ cut NdGaO$_{3}$ substrate (where subscript 'o' stands for orthorhombic unit cell in Pbnm space group with a = 5.433 \AA, b = 5.504 \AA, c= 7.716 \AA\ ). The low magnification HAADF-STEM image of both are given in Supplementary Information, Fig. S1 \& S2.  
Fig.\ref{fig:Figure 1}e\& f show the maps of o-o-p strain from GPA analysis. The SmNiO$_{3}$ films have a uniform o-o-p reduction of 3\% as compared to the substrate, corresponding to an o-o-p parameter of 3.78 \AA, while the Sm$_{9}$Ni$_{9}$O$_{22}$ films have a uniform o-o-p parameter of 3.63 \AA. Such uniformity indicates a high crystalline quality devoid of any defects, which is also visible at larger field of view (see Supplementary Information, Fig. S1). Furthermore, the Sm$_{9}$Ni$_{9}$O$_{22}$ have faint but robust contrasts corresponding to (303)$_{\text{pc}}$ (subscript pc refers to the pseudocubic unit cell) oriented stripes which result in superstructure spots in the Fourier transform (FT) shown in Fig.\ref{fig:Figure 1}h at Q = ($\pm \frac{1}{3}, 0, \pm \frac{1}{3}$), ($\pm \frac{2}{3}, 0, \pm \frac{2}{3}$), ($\pm \frac{1}{3}, 0, \mp \frac{2}{3}$), ($\pm \frac{2}{3}, 0, \mp \frac{1}{3}$) r.l.u. Here we define the reciprocal lattice units (r.l.u.), with in plane components a = b = 3.86 \AA, \ and the o-o-p lattice constant c = 3.63 \AA\ for the Sm$_{9}$Ni$_{9}$O$_{22}$.

\begin{figure*}[h!]
\centering
\includegraphics[width=\textwidth]{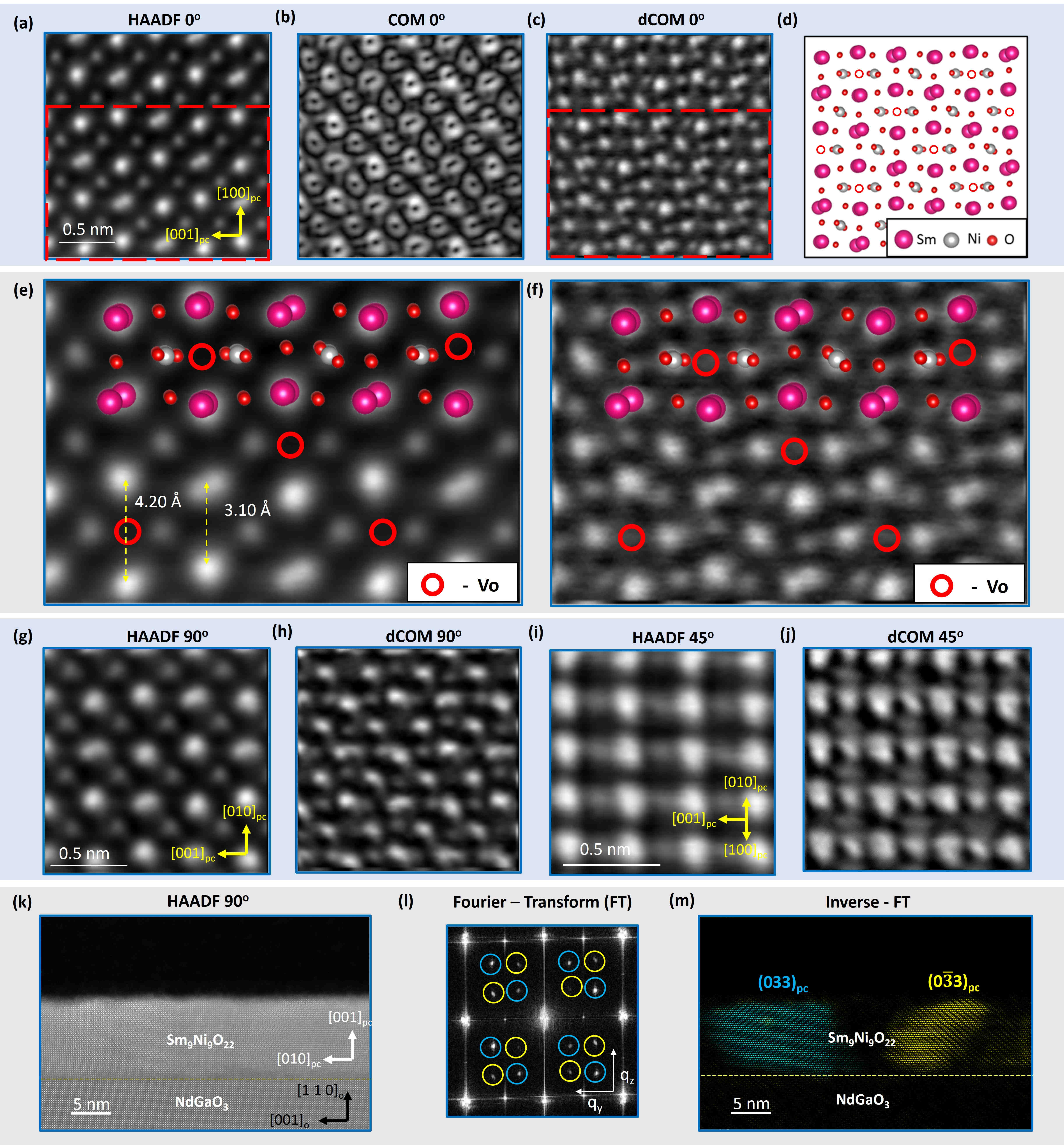}
\caption{4D-STEM analysis of Sm$_{9}$Ni$_{9}$O$_{22}$ in 0$^{\circ}$, 90$^{\circ}$, and 45$^{\circ}$ orientations that corresponds to [010], [100], and [110]$_{\text{pc}}$ viewing axis. (a) HAADF image of Sm$_{9}$Ni$_{9}$O$_{22}$ viewed along [010], showing doubled rare-earth sites and elongated - shortened Sm-Sm distances. (b) The COM image of the same region, obtained by 4D-STEM analysis. (c) The corresponding dCOM image, with a very good oxygen phase contrast, that shows Vo line in-between the extended Sm-Sm distance, hence along the stripes. (d) A relaxed structural model that is identical to the dCOM image in (c). (e) A magnified HAADF image of region highlighted in (a), indicating the location of each atom, and showing the extended and reduced Sm-Sm distances. (f) A magnified dCOM image of region highlighted in (c), indicating the location of each atom, and periodic Vo. (g) HAADF image of Sm$_{9}$Ni$_{9}$O$_{22}$ viewed along [100], showing features similar as in the view along [010]. (h) The corresponding dCOM image. (i) HAADF image of Sm$_{9}$Ni$_{9}$O$_{22}$ viewed along [110]. (j) The corresponding dCOM image. (k) A low magnification HAADF image of Sm$_{9}$Ni$_{9}$O$_{22}$ viewed along [100], that shows stripes along (033)$_{\text{pc}}$ and (0$\bar{3}$3)$_{\text{pc}}$directions. (l) The corresponding FT showing spots at \textbf{Q} = (0, $\pm \frac{1}{3},\pm \frac{1}{3}$), (0, $\pm \frac{2}{3},\pm \frac{2}{3}$), (0, $\pm \frac{1}{3},\mp \frac{2}{3}$), (0, $\pm \frac{2}{3},\mp \frac{1}{3}$) r.l.u. and \textbf{Q} = (0, $\pm \frac{1}{3},\pm \frac{2}{3}$), (0, $\pm \frac{2}{3},\pm \frac{1}{3}$), (0, $\pm \frac{1}{3},\mp \frac{1}{3}$), (0, $\pm \frac{2}{3},\mp \frac{2}{3}$) r.l.u. (m) The inverse-FT image showing the two domains obtained from the highlighted spots in (l).}
\label{fig:Figure 2}
\end{figure*}

Complementary 4D-STEM measurements were carried out to further structurally characterize this new phase (Sm$_{9}$Ni$_{9}$O$_{22}$) and determine the local structural modifications at the atomic level. Fig.\ref{fig:Figure 2} shows the HAADF and 4D-STEM results obtained along different crystallographic orientations of the Sm$_{9}$Ni$_{9}$O$_{22}$.  In Fig.\ref{fig:Figure 2}a,e are the Sm$_{9}$Ni$_{9}$O$_{22}$ HAADF images at higher magnification, corresponding to a projection of the NGO substrate along the [001]$_{\text{o}}$ zone axis. This is denoted as the 0$^{\circ}$ projection, or [010]$_{\text{pc}}$ zone axis, for the new Sm$_{9}$Ni$_{9}$O$_{22}$ phase. The HAADF image shows the extended and reduced Sm-Sm distances, and one third of the Sm column exhibit a doubling, with all these reconstructions running periodically along the \{303\}$_{\text{pc}}$ stripes.
4D-STEM data collected simultaneously with the HAADF images are shown in Fig.\ref{fig:Figure 2}b,c,f. Figure 2b corresponds to the displacement intensity of the center-of-mass (COM) of the diffraction pattern measured at each pixel, and donut shapes reveal the presence of atomic columns at their centers, even for oxygen atoms  \cite{yang2021determination,jia2023real}. The figures 2c,f are obtained by the divergence of the center-of mass (dCOM) approximating a projected charge-density image, with extrema located at the atomic positions. These images reveal unambiguously that columns of apical oxygen vacancies (Vo) are ordered along the stripes. This is causing the rare-earth super-structure, since the Sm atoms are moving away from the Vo sites giving separations between the Sm atoms of $\approx$ 4.20 \AA\ as shown in Fig.\ref{fig:Figure 2}e (while a reduced Sm-Sm distance  of $\approx$ 3.10 Å  is measured when columns of apical oxygen are  preserved). Such local variations in the A-cation distances, arising due to the coexistence of different coordinations of B sites have been observed in studies of cobaltite\cite{kim2012probing}, ferrite\cite{hirai2013anisotropic}, and manganite\cite{parsons2009synthesis} systems.

A structural model showing a similar atomic reconstruction is shown in Fig.\ref{fig:Figure 2}d. This was obtained by an ab-initio crystal relaxation of a cubic SmNiO$_{3}$, after introducing a single family of (303)$_{\text{pc}}$ planes of apical Vo. The presence of such planes of Vo leads to a mixture of octahedral and pyramidal Ni sites, and it forms the n = 3 member of a family made of (n0n)$_{\text{pc}}$ planes of Vo, with a formula A$_{n}$B$_{n}$O$_{3n-1}$. Interestingly, infinite-layer nickelates, with the formula RNiO$_{2}$ form the n = 1 member. The structural details of such A$_{n}$B$_{n}$O$_{3n-1}$ series as predicted by ab-initio calculations are in Supplementary Information, Fig.S3 and S4.

The STEM images measured along a projection rotated by 90° compared to Fig.\ref{fig:Figure 2}a (corresponding to a substrate [$\bar{1}$10]$_{\text{o}}$ zone axis and a [100]$_{\text{pc}}$ Sm$_{9}$Ni$_{9}$O$_{22}$ zone axis )  as shown in Fig.\ref{fig:Figure 2}g and h indicate that the vacancy ordering is taking place along two perpendicular faces. 
While the observation along the 0° projection of the Sm$_{9}$Ni$_{9}$O$_{22}$ was only showing (303)$_{\text{pc}}$ planes (e.g. and not a ($\bar{3}$03)$_{\text{pc}}$), the Fig.\ref{fig:Figure 2}k-m shows different domains in 90$^{\circ}$ projections, where the stripes cpincide with the (033)$_{\text{pc}}$ and the (0$\bar{3}$3)$_{\text{pc}}$ planes. This indicates that  0$^{\circ}$ and 90$^{\circ}$ projections are not exactly symmetrical, which is most likely due to the orthorhombicity of the substrate, that causes an asymmetry between the two in-plane directions. 

Nevertheless, locally both orientations reveal very similar cation and oxygen atomic reconstruction of the Sm$_{9}$Ni$_{9}$O$_{22}$. This essentially identifies that this phase is obtained by intersecting apical Vo planes of (303)$_{\text{pc}}$ and (033)$_{\text{pc}}$ orientations.     
Such a possible tetragonal symmetry is also confirmed by the observation along 45$^{\circ}$ projection as shown in Fig.\ref{fig:Figure 2}i and j. The apical oxygen removal indicatively reduces the o-o-p parameter, as is the case for infinite-layer nickelates, where the complete removal of apical oxygen results in an o-o-p parameter of $\approx$ 3.30 \AA \  \cite{li2019superconductivity, wang2020synthesis,wei2023solid,krieger2022synthesis}. The observed o-o-p parameter of 3.63 \AA \ in Sm$_{9}$Ni$_{9}$O$_{22}$ is thus associated with these oxygen deficiencies.
  
\subsection*{Structural and charge ordering from ab-initio calculations}
\begin{figure*}[h!]
\centering
\includegraphics[width=\textwidth]{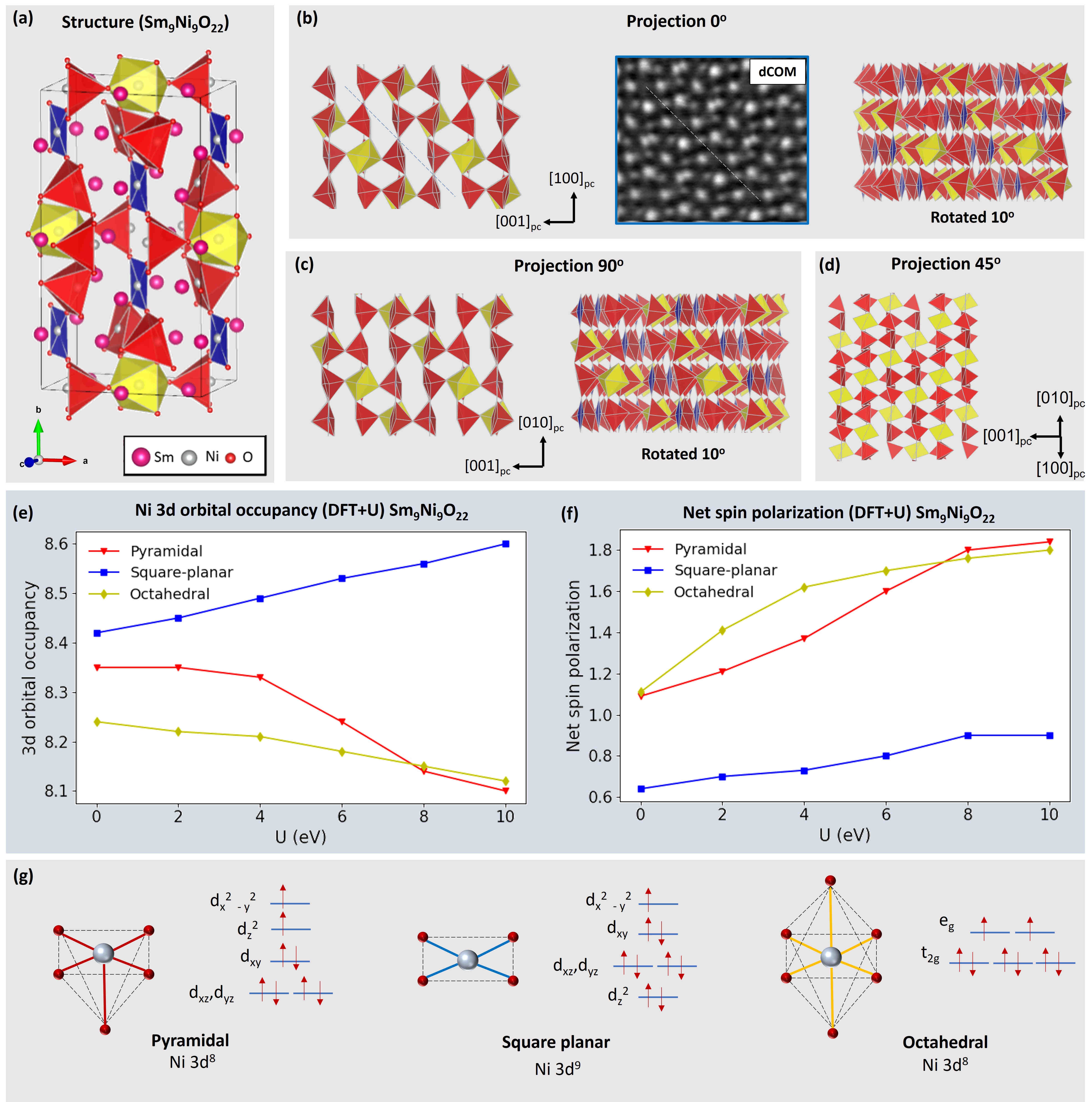}
\caption{Structural and electronic refinement by DFT. (a) Relaxed structural model of Sm$_{9}$Ni$_{9}$O$_{22}$, with Ni sites in a mixed coordination of pyramidal, square planar and octahedral. The axis indicated in this figure is the local axis with respect to the structure symmetry. (b) The polyhedral distribution viewed along [010]$_{\text{pc}}$ and a 10$^{\circ}$ rotation along [100], with the area matching with that of the dCOM image. (c) The polyhedral distribution viewed along [100]$_{\text{pc}}$ and a 10$^{\circ}$ rotation along [010]. (d) The polyhedral distribution viewed along [110]. From viewing along the three projections, the different polyhedra appear to have a periodic distribution. (e) Calculated 3d orbital occupancy of different Ni sites for U = 0 to 10 eV by DFT+U. (f) Calculated net spin polarization of different Ni sites for U = 0 to 10 eV by DFT+U. (g) A demonstration of the d-orbital occupancy of different Ni sites. Ni in the pyramidal and octahedral sites are in a 3d$^{8}$ configuration, while Ni in the square planar sites are in a 3d$^{9}$ configuration. }
\label{fig:Figure 3}
\end{figure*}

From these microscopic results, it is evident that this new phase cannot be simply explained by one family of Vo planes, and an intersecting (303)$_{\text{pc}}$ and (033)$_{\text{pc}}$ planes of apical vacancies are needed. Corresponding simulations involved the structural relaxations of a 6x6x6 cubic SmNiO$_{3}$ supercell with (303)$_{\text{pc}}$ and (033)$_{\text{pc}}$ ordered apical Vo. The calculations resulted in a pseudo-tetragonal structure with an o-o-p parameter of $c = 3.63$ \AA. This 6x6x6 tetragonal supercell was then symmetrized, giving an 80-atom structure with the formula Sm$_{18}$Ni$_{18}$O$_{44}$, as shown in Fig.\ref{fig:Figure 3}a, which can be further shortened as Sm$_{9}$Ni$_{9}$O$_{22}$. More details are given in Methods and Supplementary Information, Fig S6 and table S1. This structure essentially identifies three types of Ni sites: NiO$_{6}$ octahedral (Oh), NiO$_{4}$ square-planar (Sq) and NiO$_{5}$ pyramidal (Py) in the ratio 1:2:6 in the lattice. In this regard, it is very different from the A$_{3}$B$_{3}$O$_{7}$ \cite{moriga1994reduction}, which have the Vo in the periodicity $\frac{1}{3}$ from the parent ABO$_{3}$ structure  and is built with apical Vo lines running only along one face. It is here also noted that  an alternative A$_{3}$B$_{3}$O$_{7}$ structure can also been obtained by mixing apical Vo (303)$_{\text{pc}}$ planes and  equatorial Vo (033)$_{\text{pc}}$ planes, but the resulting structure would not match our observation (See Supplementary Information, Fig. S5).

On the grounds of differently coordinated Ni sites in the lattice, a mixed valence is expected for the Sm$_{9}$Ni$_{9}$O$_{22}$ composition. Ab-initio calculations give evidence of different Ni electronic configurations for the different site coordination as indicated by Fig.\ref{fig:Figure 3}e-g. Calculations have been done including a Hubbard U parameter ranging from 0 to 10 eV as shown in Fig.\ref{fig:Figure 3}e,f. At U = 0 eV, differences of total charges on the Ni sites are already present, but become more segregated between the square and the octahedra or the pyramid when U increases. The Ni in square planar sites are in a 3d$^{9}$ configuration, indicating Ni$^{1+}$ and a spin = 1/2 (typical occupation matrix is given in Supplementary Information, Table S2). The Ni in the pyramidal and octahedral sites are closer to a 3d$^{8}$ configuration and a spin = 1. 
A Sm$_{9}$Ni$_{9}$O$_{22}$ composition would imply 17 charges to be distributed over the nickel sites and a possible distribution would then be 2 sq(Ni$^{+}$, d$^{9}$), 6 Py(Ni$^{2+}$, d$^{8}$) and a single Oh (Ni$^{3+}$, 3d$^{8}$$\underline L$ with $\underline L$ referring to a hole in the ligand).
The exact nature of the Ni$^{3+}$ ground state (i.e. the respective weight of 3$d^{7}$ and 3$d^{8}\underline L$ in the ground state) is difficult to address solely based from the DFT+U level where, in brief, electron localization and correlation are brutally parameterized. 
We also assume that a ligand hole will reside on the octahedral sites by analogy from the electronic structure of the parent SmNiO$_{3}$ phase.
In the metallic state, it is known to host octahedral Ni$^{3+}$ with primary covalent ground state  (a.k.a, as negative charge transfer), id est, primarily composed by a 3d$^{8}$$\underline L$. In the insulating state, the bond disproportionates with the presence of collapsed $d^{8}\underline L^{2}$ and expanded $d^{8}$ octahedral sites \cite{green2016bond}. These charge distribution and orbital physics are summarized in figure  \ref{fig:Figure 3}g, and they are further probed by spectroscopic techniques involving x-ray absorption spectroscopy (XAS), resonant inelastix x-ray scattering (RIXS) and hard x-ray photoemission spectroscopy (HAXPES).   
 
\subsection*{Orbital polarization and charge localization by spectroscopic measurements}
\begin{figure*}[h!]
\centering
\includegraphics[width=\textwidth]{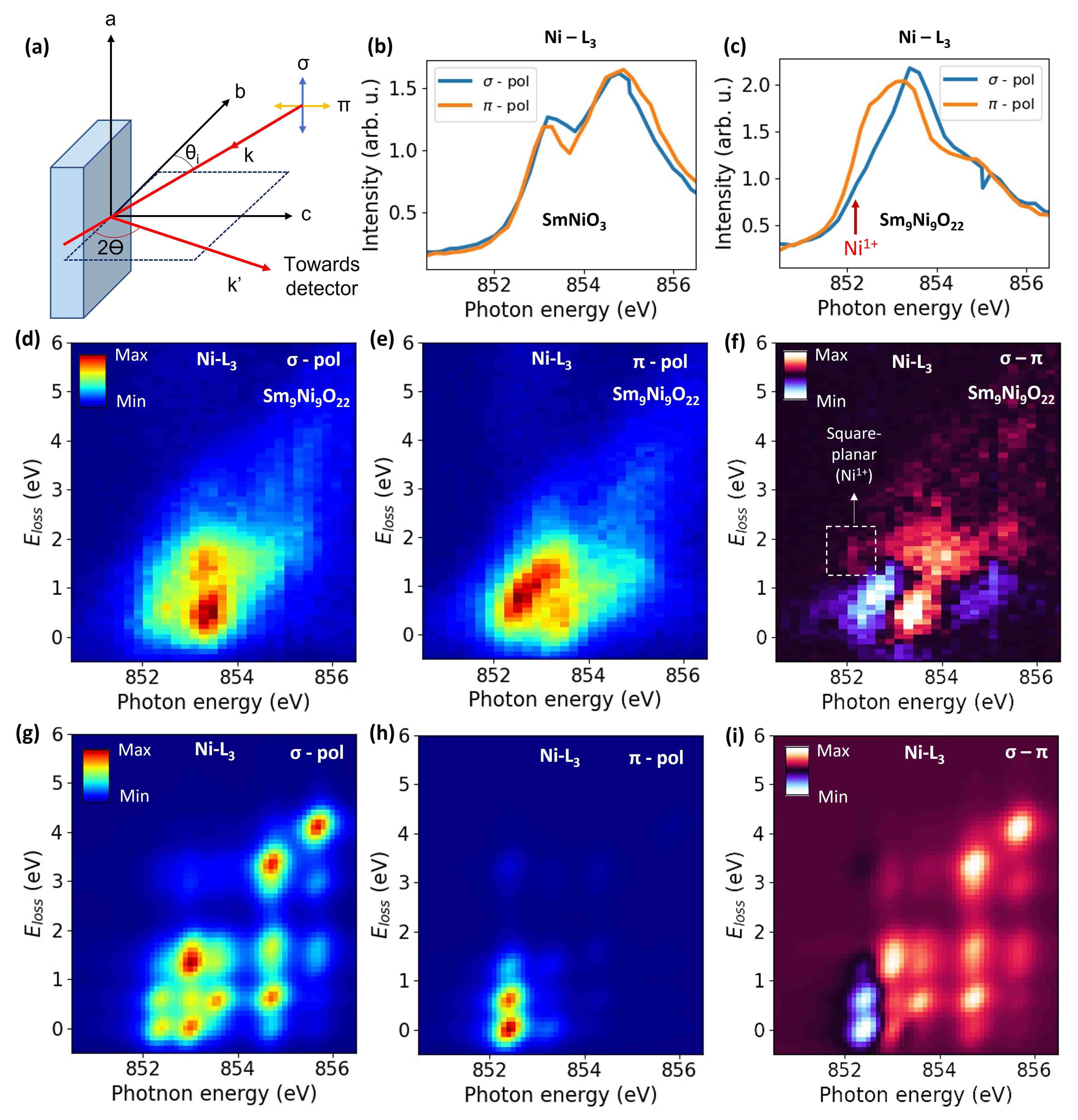}
\caption{Electronic structure analysis by XAS and RIXS. (a) The experimental geometry showing the incident x-ray polarization $\sigma$ and $\pi$, incident x-ray angle $\theta_{i}$ and the scattering angle 2$\Theta$.(b \& c) XAS partial fluorescence yield (PFY) spectra at 25 K at Ni-L$_{3}$ edge of (b) SmNiO$_{3}$ (c) Sm$_{9}$Ni$_{9}$O$_{22}$. (d \& e) Ni-L$_{3}$ edge RIXS map at 25 K of Sm$_{9}$Ni$_{9}$O$_{22}$ in (d) $\sigma$ - incident polarization and (e) $\pi$ - incident polarization of the photons. (f) Difference of RIXS map between the two polarizations. (g \& h) Calculated Ni-L$_{3}$ edge RIXS map of Ni$^{2+}$ pyramidal sites, in (g) $\sigma$ polarization and (h) $\pi$ polarization.(i) Difference of calculated RIXS map between the two polarizations.}
\label{fig:Figure 4}
\end{figure*}

The schematic of the XAS and RIXS measurements is briefly indicated in Fig.\ref{fig:Figure 4}a. 
The XAS Ni-L$_{3}$ edge of the parent perovskite SmNiO$_{3}$ thin film measured at 25K is given in Fig.\ref{fig:Figure 4}b which shows a typical splitting associated to bond disproportionation and the presence of collapsed $d^{8}\underline L^{2}$ and expanded $d^{8}$  octahedra for the insulating state \cite{green2016bond}. A more complete set of the XAS and RIXS data at the Ni-L edges measured for the parent SmNiO$_{3}$ thin film at RT (300K) and 25K can be seen in Supplementary Information, Fig.S7. They are in agreement with previously reported spectroscopic work on perovskite nickelate thin films \cite{bisogni2016ground}, and they show a fully developed insulating behavior at 25K, and a partial metal-to-insulator transition occurring at RT (300K) as revealed by a blurring of the XAS Ni-L$_{3}$ edges together with a blue shift of the localized dd-orbital excitations in the RIXS \cite{bisogni2016ground}. 
The Ni-L$_{3}$ edge XAS for the Sm$_{9}$Ni$_{9}$O$_{22}$ sample (Fig.\ref{fig:Figure 4}c) occurs at a lower excitation energy than that of SmNiO$_{3}$ indicating a reduced valence toward Ni$^{2+}$ and Ni$^{1+}$ as it was inferred from the structural investigation. 
In complement, a Ni 2p$_{\frac{3}{2}}$ and Ni 2p$_{\frac{1}{2}}$ core level bulk sensitive HAXPES measurements between Sm$_{9}$Ni$_{9}$O$_{22}$ and SmNiO$_{3}$ are given in (Supplementary Information, Fig. S13) that show a similar valence trend. 
In the XAS, a strong dichroism is observed at the Ni-L edges of the Sm$_{9}$Ni$_{9}$O$_{22}$ with a noticeable o-o-p higher-intensity at the low-energy side of the Ni-L$_{3}$ main edge (at 852.5 eV). From atomic multiplet calculations (shown in Supplementary Information, Fig.S14) \cite{stavitski2010ctm4xas}, this strong dichroism is considered to be stemming from the Ni$^{2+}$ in the pyramidal sites, which are the most populated in the lattice. 
By projecting the XAS final-state 2p$^{5}$3d$^{9}$ in terms of crystal field configuration,  the features at 852.5 eV and 853.3 eV are identified respectively with 3d$_{z^{2} - r^{2}}$ and 3d$_{x^{2} - y^{2}}$ symmetry of the pyramidal sites explaining the sign and the hierarchy of the XLD signal \cite{stavitski2010ctm4xas,ikeno2009multiplet}.  An in-plane contribution is also expected from the XAS spectra of the square planar 3d$_{x^{2} - y^{2}}$ state of the Ni$^{1+}$ and it is at the origin of the shoulder observed at 852.3 eV for the  $\sigma$  polarization. Nevertheless, the overall shape of the XAS-XLD is dominated by the Ni$^{2+}$ in the pyramidal site with only a smaller contribution of the Ni$^{1+}$ site in the square plane.
These multi-valent features of the Sm$_{9}$Ni$_{9}$O$_{22}$ structure are made clearer in the RIXS maps. The data for the $\sigma$ and $\pi$ polarization, along with their linear dichroism are displayed in Fig.\ref{fig:Figure 4} d-f. The RIXS data show a similar dichroism as observed in XAS, e.g. with a larger contribution in the $\pi$ polarization at $\approx$ 852.5 eV. The calculated RIXS maps obtained for a Ni$^{2+}$ within a pyramid is in a fair agreement with the experimental maps (Fig.\ref{fig:Figure 4}g-i). A noticeable difference on comparing Fig\ref{fig:Figure 4}f and i is the experimental intensity at an energy loss of 1.7 eV obtained for an excitation energy of $\approx$ 852.3 eV, that is not present in the calculation done for Ni$^{2+}$. This excitation energy corresponds to Ni$^{1+}$ and its higher intensity for a $\sigma$ polarization is consistent with a square planar symmetry (see calculation in figure S15), in accordance with 3d$_{x^{2} - y^{2}}$ $\longrightarrow$ 3d$_{xy}$, 3d$_{xz}$/3d$_{yz}$ d-d transitions as also discussed by Rossi et al. \cite{rossi2021orbital} in the case of an IL-nickelate. The RIXS signatures have also been investigated at RT for some excitation energies. The dichroic nature of Sm$_{9}$Ni$_{9}$O$_{22}$ is rather similar at room temperature, as shown in (Supplementary Information, Fig. S7 and S8) but some d-d transitions experience blue shift of $\approx$ 0.2 eV indicating electronic modification between low to room temperature.
\begin{figure*}[h!]
\centering
\includegraphics[width=\textwidth]{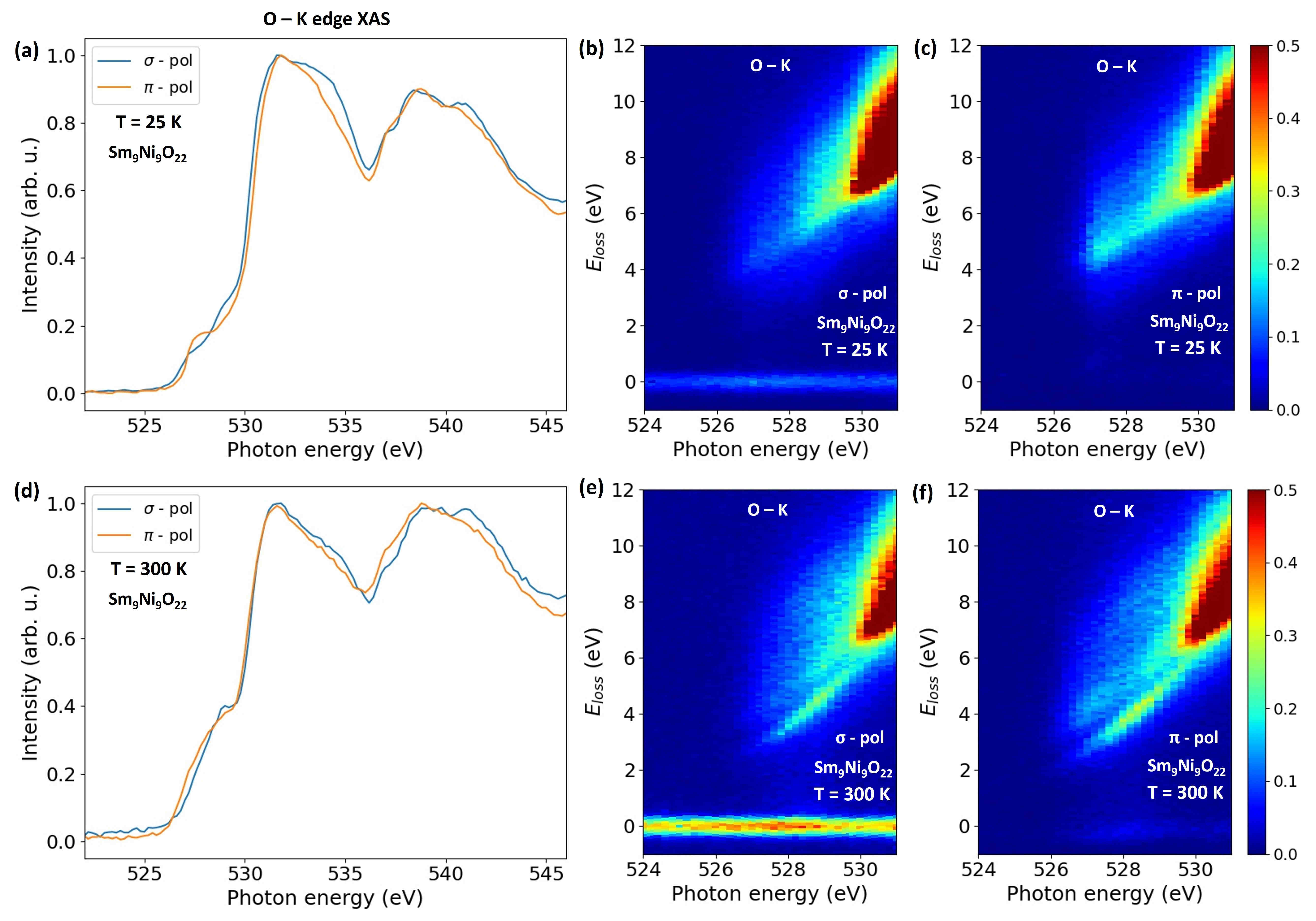}
\caption{Temperature dependent spectroscopic studies at the O-K edge by XAS and RIXS maps (in energy loss) of Sm$_{9}$Ni$_{9}$O$_{22}$. (a-c) Measurements at T = 25 K, (a) O-K edge PFY XAS spectra in $\sigma$ and $\pi$ incident polarizations, (b \& c) RIXS map at O-K edge for (b) $\sigma$ incident photon polarization and (c) $\pi$ incident photon polarization. (d-f) Measurements at T = 300 K, (d) O-K edge PFY XAS spectra in $\sigma$ and $\pi$ incident polarizations, (e \& f) RIXS map at O-K edge for (e) $\sigma$ incident photon polarization and (f) $\pi$ incident photon polarization.}
\label{fig:Figure 5}
\end{figure*}

In complement, similar spectroscopic studies have been carried out at the O-K edge for the parent and reduced phase, both at low and room temperature. O-K edge XAS has been previously used to investigate the electronic transition in a perovskite nickelate \cite{meyers2016pure}. Here in the parent phase, along with the XAS, the electronic transition is further characterized by the RIXS maps that also indicates a band gap opening at low temperature (Supplementary Information, Figure S9 and S10).

The O-K edge XAS and RIXS maps of Sm$_{9}$Ni$_{9}$O$_{22}$ shown in Fig.\ref{fig:Figure 5}a-f and in Supplementary Information, Fig.S11 \& S12 are very different from the parent SmNiO$_{3}$. The pre-peak of the O-K is very weak which also confirms a reduced valency and co-valency as compared to the Ni$^{3+}$ of SmNiO$_{3}$. The position of the pre-peak also confirms a band gap opening of $\approx$ 3 eV for the reduced Sm$_{9}$Ni$_{9}$O$_{22}$. While being of small intensity, the pre-peak shows a strong incident photon polarization dependence. At low temperature, this pre-peak intensity is successively stronger in energy for a $\sigma$, a $\pi$ and then a $\sigma$ polarization that could be interpreted by a combination of hybridization of the O-2p orbital with the unoccupied 3d$_{x^{2}-y^{2}}$ of the Ni$^{1+}$ and then with the 3d$_{z^{2}}$ and 3d$_{x^{2}-y^{2}}$ of the Ni$^{2+}$. This goes hand in hand with the respective  energy positions of the calculated unoccupied density of states of these orbitals as shown in (Supplementary Information, Fig.S16). 

The XAS O-K pre-edge at room temperature seems smeared out, and exhibits a weaker dichroism. While observing the RIXS maps, a strong fluorescence line appears at room temperature causing the blurring of the O-K XAS pre-edge feature. This fluorescence line occurs for an incident photon energy corresponding to Ni-O hybridization and with the smallest energy loss, involving energy bands at the upper part of the valence band. A disappearance with temperature of this fluorescence line reveals an electronic phase transition and somehow a charge localization at low temperature for the Sm$_{9}$Ni$_{9}$O$_{22}$. Due to the weak dichroic character of this band, the localisation might concern the Ni$^{3+}$ in the octahedra whose electron configuration in the parent phase is also prone to charge localization \cite{mikheev2015tuning}.  Futhermore, additional electrons associated to oxygen deficient perovskite nickelates have also been predicted to strongly localize on octahedral sites, explaining the dramatic decrease in the electrical conductivity by several orders of magnitude in them \cite{kotiuga2019carrier}.

\section*{Discussion}\label{sec11}
Here, the topotactic reduction of a parent perovskite SmNiO$_{3}$ thin-film grown on NdGaO$_{3}$ results in a homogeneous thin-film of Sm$_{9}$Ni$_{9}$O$_{22}$ phase, with a tetragonal symmetry and a reduced o-o-p parameter of 3.63 \AA. This structure emerges from families of \{303\}$_{\text{pc}}$ apical oxygen vacancy (Vo) ordering giving a formula A$_{9}$B$_{9}$O$_{22}$ as revealed by 4D-STEM.
A single family plane will give A$_{3}$B$_{3}$O$_{8}$ phase with only octahedral and pyramidal Ni sites. Here we show that the A$_{9}$B$_{9}$O$_{22}$ phase emerges from Vo families along two orientations, giving rise to square-planar, pyramidal and octahedral Ni sites.
Square-planar environment occurs at the intersection of orthogonal families and harbours a Ni$^{1+}$ indicating a possible route to IL-structure.
It is to note that such (303)$_{\text{pc}}$ phases have also been reported as structural defects, constrained in quasi-2D nano-sheet on the top of IL-NNO2 \cite{raji2023charge}. In fact, such (303)$_{\text{pc}}$ defect might be present as nanodomains, presumably along a single orientation, or overlap along different orientation and order at longer range as we showed here, enabling phase engineering in the ABO$_{2}$-ABO$_{3}$ system. In such case, this A$_{9}$B$_{9}$O$_{22}$ phase could be a member of a potentially new family of nickel oxides, that forms by \{n0n\}$_{\text{pc}}$ ordered Vo. Overlapping (n0n)$_{\text{pc}}$ and (0nn)$_{\text{pc}}$ apical oxygen vacancy planes results in a general formula A$_{n^{2}}$B$_{n^{2}}$O$_{3n^{2}-2n+1}$ that is indeed spanning from the IL-ABO2 with only square planar to the perovskite ABO$_{3}$ with only octahedral Ni sites. In-between, octahedral, square-planar and pyramidal sites are in the ratio $(n-2)^{2}:2:(4n-6)$. Such \{n0n\}$_{\text{pc}}$ planes could be seen as an analogous case of the Ruddlesden-Popper (RP) faults that is also present from the individual 2D defects and scale to a fully developed RP phase.  

The stability of this family is still to be understood and being an intermediate phase, it will depend on epitaxial strain and redox condition. This is in line with our observations indicating that the orthorhombic substrate plays a role, as strain in thin films is generally known to control the creation of Vo, even for the stabilization of superconducting IL-nickelate films \cite{lee2023linear}. Additionally, the choice of the reducing agent such as CaH$_{2}$ \cite{li2019superconductivity}, alumina \cite{wei2023superconducting} and atomic hydrogen \cite{lee2023linear} can influence the selective stabilization of each phase. The A$_{9}$B$_{9}$O$_{22}$ phase is also a neighbour of other members of the oxygen deficient nickelate series, such as the brownmillerite ABO$_{2.5}$ \cite{wu2023topotactically} and the A$_{3}$B$_{3}$O$_{7}$ \cite{moriga1994reduction}. 
The chemical dynamics behind the selective stabilization of these phases is still not understood. 
Possibly, this new phase A$_{9}$B$_{9}$O$_{22}$ might have been present along the samples that reported the A$_{3}$B$_{3}$O$_{7}$ \cite{wang2020synthesis,wei2023solid}, or it might have been misinterpreted as that. 

Interestingly, ab-initio calculations and spectroscopic studies coherently indicate that the Sm$_{9}$Ni$_{9}$O$_{22}$ phase consists of tri-component Ni sites, with a varying valence from Ni$^{1+}$ to Ni$^{3+}$. Each component is associated with a unique symmetry, charge, orbital, and spin distribution, that collectively express a multivalent character with different degrees of orbital polarisation at each site, as evidence by our spectroscopic studies. The tri-components are distributed in the lattice with a periodicity determined by the Vo ordering, contributing to a commensurate charge order for this system with a periodicity ($\frac{1}{3}$, 0, $\frac{1}{3}$) r.l.u. This periodicity is strikingly similar with that of the CO reported in IL-nickelate thin films \cite{tam2022charge,ren2023symmetry, krieger2022charge,rossi2022broken}. This possibly indicates the existence of this phase in IL-nickelate thin-films as a defect, and this is further strengthened from recent studies reporting apical oxygen ordering in IL-nickelate thin-films that exhibited similar CO \cite{raji2023charge,parzyck2023absence}. It questions how intrinsic is the ($\frac{1}{3}$, 0) r.l.u. CO in IL-NdNiO$_{2}$. 

The parent perovskite SmNiO$_{3}$ exhibits a metal-to-insulator (MIT) around 375K as evidenced by our transport measurements, and this is similar to what was previously reported \cite{torriss2017metal} in these. However the novel system 
Sm$_{9}$Ni$_{9}$O$_{22}$ has a complex multivalent electronic landscape, with an intriguing carrier localization contributing to an enhanced resistivity. This is in contrast with the case of oxide semiconductors, where Vo cause an enhanced conductivity. However, this is in line with previous reports on perovskite nickelates, where the Vo cause localization of electrons at the octahedral Ni sites \cite{kotiuga2019carrier}, with the subsequent filling of the ligand hole resulting in a 3$d^{8}$ configuration. The oxygen deficient nickelate discussed in that report might correspond to this novel phase. Furthermore, RIXS evidence an additional charge localization at low temperature, presumably associated to the charge configuration at octahedral Ni site. It is worth noting that the Ni sites at the parent phase octahedra have known charge fluctuation 3$d^{8}\underline L$ $\longleftrightarrow$ 3$d^{8}$ + 3$d^{8}\underline L^{2}$ playing a role in the temperature driven MIT. This might be hindered for an isolated octahedron, giving rise to different charge fluctuation (e.g. 3$d^{8}\underline L$ $\longleftrightarrow$ 3$d^{7}$), making the phase diagram of this system more complex. 

On a broader perspective, this is another example of the idea of selective oxygen incorporation driving stabilization of phases/interfaces that have metallic, semiconducting, insulating, and even superconducting (with doping) properties \cite{li2019superconductivity, chen2018roles, suntivich2011design, liu2014dominant}. The carrier distribution in the lattice, which is associated to the distribution of oxygen atoms, has an identical effect as that of hydrogenated perovskite nickelate devices proposed for neuromorphic computing applications \cite{zhang2022reconfigurable}. This was demonstrated for an oxygen deficient NdNiO$_{3-\delta}$ nickelate device, that could potentially have this A$_{9}$B$_{9}$O$_{22}$ phase, where multiple resistance states were achieved by adjusting the electronic bias \cite{kotiuga2019carrier}. Such findings uncover the possible multitude of applications of this phase on par with other strongly correlated electron systems.      

\section*{Acknowledgements}
A.G., A.N., and J.-P.R. acknowledges financing from EU-H2020 IMPRESS under grant agreement (n. 101094299). A.R. acknowledges financing from LABEX NanoSaclay and EU-H2020 ESTEEM-3 under grant agreement (n. 823717) for the doctoral funding. A.G. and X.L. also acknowledges financing from EU-H2020 ESTEEM-3 under grant agreement (n. 823717).  Nion UltraSTEM at LPS Orsay and the FIB at C2N, University of Paris-Saclay was accessed in the TEMPOS project framework (ANR 10-EQPX-0050). Z.D., B.F. and D.L. acknowledge the funding support from the Natural Science Foundation of China (No. 12174325) and Guang Dong Basic and Applied Basic Research Foundation (2023A1515011352). The work performed in Hong Kong is supported by the Research Grants Council of Hong Kong through the Early Career Scheme (CityU 21301221) and the General Research Fund (CityU 11309622). L.V., C.D. and M.H. acknowledge support by the Swiss National Science Foundation - division II (grant no. 200020\_179155 and 200020\_207338), and by the European Research Council under the European Union Seventh Framework Programme (FP7/2007–2013)/ERC Grant Agreement no. 319286 (Q-MAC). We acknowledge SOLEIL Synchrotron for provision of beamtime under the proposals 20220630 and 20221574. We thank Daniele Preziosi, Alberto Zobelli, Zhizhong Li, Nicolas Jaouen and Jean-Marc Triscone for fruitful discussions.
\section*{Methods}\label{sec13}
\subsection*{Samples}
Before growth, the (110)$_{\text{o}}$-oriented NdGaO$_{3}$ single-crystalline substrates were annealed at 1150 $^\circ $C for 10 hours in flowing oxygen to obtain sharp step-and-terrace surface morphology. Then the pristine SmNiO$_3$ epitaxial thin films were grown by off-axis radio-frequency magnetron sputtering, at a substrate temperature of 460 $^\circ $C and a total pressure of 0.18 millibars with an oxygen/argon mixture of 1:3 \cite{catalano2014electronic}. After growth, the samples were cut into pieces with the size of 5$\times$2.5 mm$^2$ and loosely wrapped in the aluminum foil. The wrapped samples were then sealed under vacuum in Pyrex glass tubes, embedded with CaH$_2$ powder of $\approx$ 0.1 g. The sealed glass tubes were heated in a tube furnace to 230-240 $^\circ $C for 1h and cooled down to room temperature at a rate of 10 $^\circ $C/min.
\subsection*{High-resolution HAADF-STEM Imaging and 4D-STEM}
Cross sectional transition electron microscopy (TEM) lamellae were prepared using a focused ion beam (FIB) technique (at C2N, University of Paris-Saclay, France). Before FIB lamellae preparation, around 20 nm of amorphous carbon was deposited on top for protection. The HAADF imaging and 4D-STEM was carried out in a NION UltraSTEM 200 C3/C5-corrected scanning transmission electron microscope (STEM). The experiments were done at 200 keV with a probe current of approximately 10 pA and convergence semi-angles of 30 mrad. A MerlinEM (Quantum Detectors Ltd) in a 4×1 configuration (1024 × 256) has been installed on a Gatan ENFINA spectrometer mounted on the microscope \cite{tence2020electron}. The EELS spectra are obtained using the full 4 × 1 configuration and the 4D-STEM by selecting only one of the chips (256 × 256 pixels). For 4D-STEM, the EELS spectrometer is set into non-energy dispersive trajectories and we have used 6-bit detector mode that gives a diffraction pattern with a good signal to noise ratio without compromising much on the scanning speed. 
\subsection*{Ab-initio Calculations}
The first principles calculation were performed using the density functional theory \cite{kohn1965self} as implemented in the Quantum ESPRESSO package \cite{giannozzi2009quantum,giannozzi2017advanced}. The exchange-correlation functional was approximated by the generalized gradient approach \cite{perdew1996generalized}. Planewave cutoffs of 40 and 400 Ry were used basis-set and charge density expansions, respectively. Structural relaxations were done without considering a Hubbard U parameter, while the electronic structure were calculated with it. All DFT+U calculations were performed with a J = 1.2 eV, making U$_{effective}$ = U - J. All structural relaxations begin with creating a supercell from a cubic SmNiO$_{3}$, with $a = 3.80$ \AA. Then, the oxygen atoms along the corresponding family of Vo planes were removed.  
For the Sm$_{9}$Ni$_{9}$O$_{22}$ phase, after removing the two orthogonal families of apical Vo, it results in a 960-atom $6\times6\times6$ supercell.  The resulting pseudo-tetragonal o-o-p parameter of $c = 3.63$ \AA\ was obtained by minimizing the total force in a 6x6x6 supercells (the in-plane pseudo-tetragonal parameter $a = b = 3.90$ \AA, was frozen). This tetragonal supercell was then reduced using symmetry to an monoclinic 80-atom cell.  Thus obtained structure has the space group $C2/m$ and lattice parameters of $a = 9.02$, $b = 16.09$, $c = 6.47$ \AA\ and $\beta = 87.57^\circ$. The structural parameters with atomic positions are given in Table S1. The lattice vectors of the tetragonal 960-atom phase ($\Vec{a_{T}},\ \Vec{b_{T}},\ \Vec{c_{T}}$) expressed in the symmetrized monoclinic parameters ($\Vec{a_{m}},\ \Vec{b_{m}},\ \Vec{c_{m}}$) is: $\Vec{a_{T}} = \Vec{a_{m}} + \Vec{b_{m}} + 2\Vec{c_{m}},\ \Vec{b_{T}} = \Vec{a_{m}} - \Vec{b_{m}} + 2\Vec{c_{m}},\ \Vec{c_{T}} = 2\Vec{a_{m}} - 2\Vec{c_{m}}$.  
\subsection*{XAS and RIXS Measurements}
The x-ray absorption spectroscopy (XAS) and resonant inelastic x-ray scattering (RIXS) measurements of the Ni-L edges and the O-K edge were performed at the AERHA instrument of the SEXTANTS beamline at the SOLEIL synchrotron. A detailed optical scheme of the spectrometer that involves dispersing the x-ray photons as a function of their energy onto a two-dimensional detector is given in \cite{chiuzbuaian2014design}, and the experimental geometry in Fig.\ref{fig:Figure 4}a. The measurements were performed with the a and c sample axes in the horizontal scattering plane. Measurements at the Ni-L and O-K edges were taken with $\theta_{i}$=20$^{\circ}$ (where $\theta_{i}$ is the angle between the incoming x-ray and the sample surface), and linear dichroism was obtained by switching the incident x-ray polarization between $\sigma$ and $\pi$. RIXS was measured at 2$\Theta$=85$^{\circ}$, with an overall full-width at half-maximum (FWHM) energy resolution of 170$\pm$10 meV for the O-K edge and 300$\pm$10 meV  for the Ni-L edge. The measurements were done at T = 25K and T = 300K. The two-dimensional RIXS data are binned in the isoenergetic direction to form spectra and the pixel to energy conversion is performed by measuring the position of the elastic line of the spectrometer while changing the beamline energy.
\subsection*{HAXPES Measurements}
The measurements were carried out at the GALAXIES beamline at the SOLEIL synchrotron \cite{rueff2015galaxies} on the HAXPES endstation \cite{ceolin2013hard} using a photon energy of 3000 eV, with an incidence angle of 10$^\circ$. The bulk sensitivity is defined from the SESSA simulations \cite{smekal2005simulation}, that gives a probing depth of around 10 nm for 10$^\circ$ incidence. About 95\% of the detected signal will be from the elements within these estimated probing depth. The synchrotron operated with a ring current of 450 mA, giving an intensity of 3.4 × 10$^{13}$ photons/s at 3000 eV, which was then reduced using a built-in filter to 5\% of the original intensity. The photoelectrons were detected using a SCIENTA Omicron EW4000 HAXPES hemispherical analyzer, and a Shirley background \cite{shirley1972high} was removed prior to fitting the core levels spectra.
\bibliography{sn-bibliography}
\appendix


\section*{Supplementary material}

Supplementary Information for

# Emergent electronic landscapes in a novel valence-ordered nickelate with tri-component nickel coordination

Aravind Raji[1,2,9], Zhengang Dong[3,4,9], Victor Porée[2], Alaska Subedi[5], Xiaoyan Li[1], Bernat Mundet[6,7], Lucia Varbaro[6], Claribel Domínguez[6], Marios Hadjimichael[6], Bohan Feng[3,4], Alessandro Nicolaou[2], Jean-Pascal Rueff[2], Danfeng Li[3,4,⋆], and Alexandre Gloter[1,⋆]

[1]Laboratoire de Physique des Solides, CNRS, Université Paris-Saclay, Orsay, 91400, France
[2]Synchrotron SOLEIL, L'Orme des Merisiers, BP 48 St Aubin, Gif sur Yvette, 91192, France
[3]Department of Physics, City University of Hong Kong, Kowloon, Hong Kong
[4]City University of Hong Kong Shenzhen Research Institute, Shenzhen, Guangdong, 518057, China
[5]CPHT, Ecole Polytechnique, Palaiseau cedex, 91128, France
[6]Department of Quantum Matter Physics, University of Geneva, Geneva, Switzerland
[7]Electron Spectrometry and Microscopy Laboratory (LSME), Institute of Physics (IPHYS), Ecole Polytechnique Fédérale de Lausanne (EPFL), Lausanne, Switzerland
[8]LCPMR, Sorbonne Université, CNRS, Paris, 75005, France
[9]These authors contributed equally to this work
⋆Corresponding authors. Emails: danfeng.li@cityu.edu.hk,
alexandre.gloter@universite-paris-saclay.fr

## Contents

# 1 Structural - Experiment

Here we present the HAADF-STEM images of the reduced $Sm_9Ni_9O_{22}$ thin-film (Fig. S1) and parent $SmNiO_3$ (Fig. S2) in a lower field of view. The observed uniform out-of-plane compression in Fig.S1b can be ascribed to the proper stabilization of the thin-film of the reduced phase, devoid of any defects. All the measurements including 4D-STEM were carried out at room temperature at 200 keV with a probe current of approximately 10 pA, in these conditions, the samples did not show any beam damage. Optimized parameters were used for data acquisition that involved enhancing the signal to noise ratio, without compromising the scanning speed.

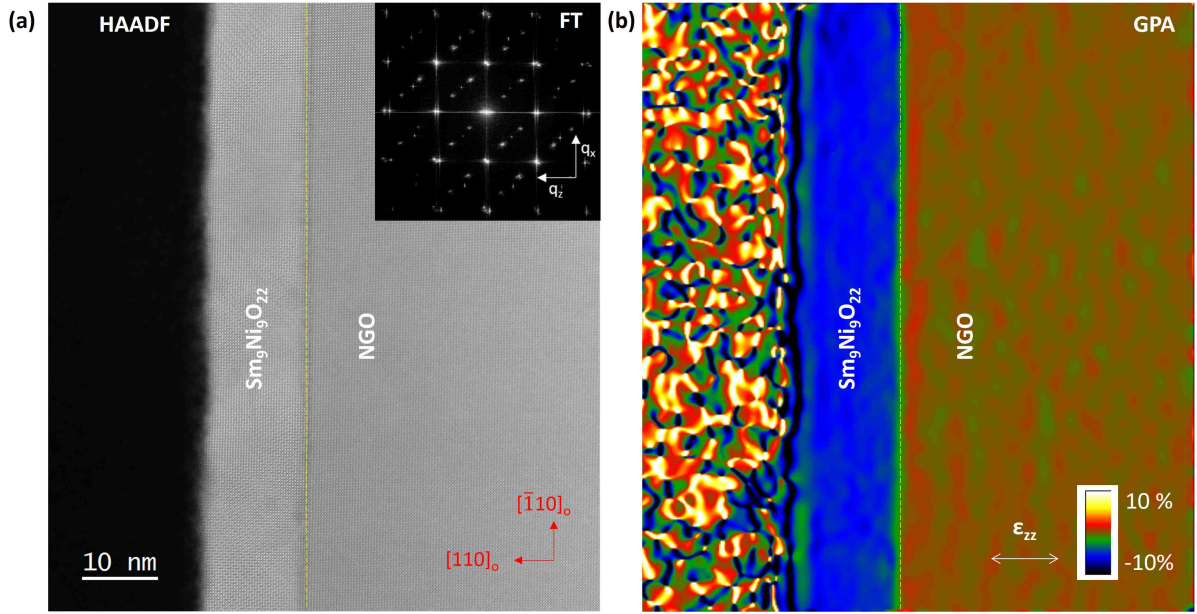

**Figure S1:** Structural homogeneity of $Sm_9Ni_9O_{22}$ thin-film. (a) Low magnification HAADF image of the $Sm_9Ni_9O_{22}$ thin-film, viewed along the $[010]_{pc}$ zone axis. A Fourier transform is shown in the inset. (b) The corresponding out-of-plane strain map by GPA. A uniform compression of 7% for the o-o-p parameter is observed for the whole $Sm_9Ni_9O_{22}$ thin-film (the substrate o-o-p is the reference), indicating the homogeneity of the thin-film.

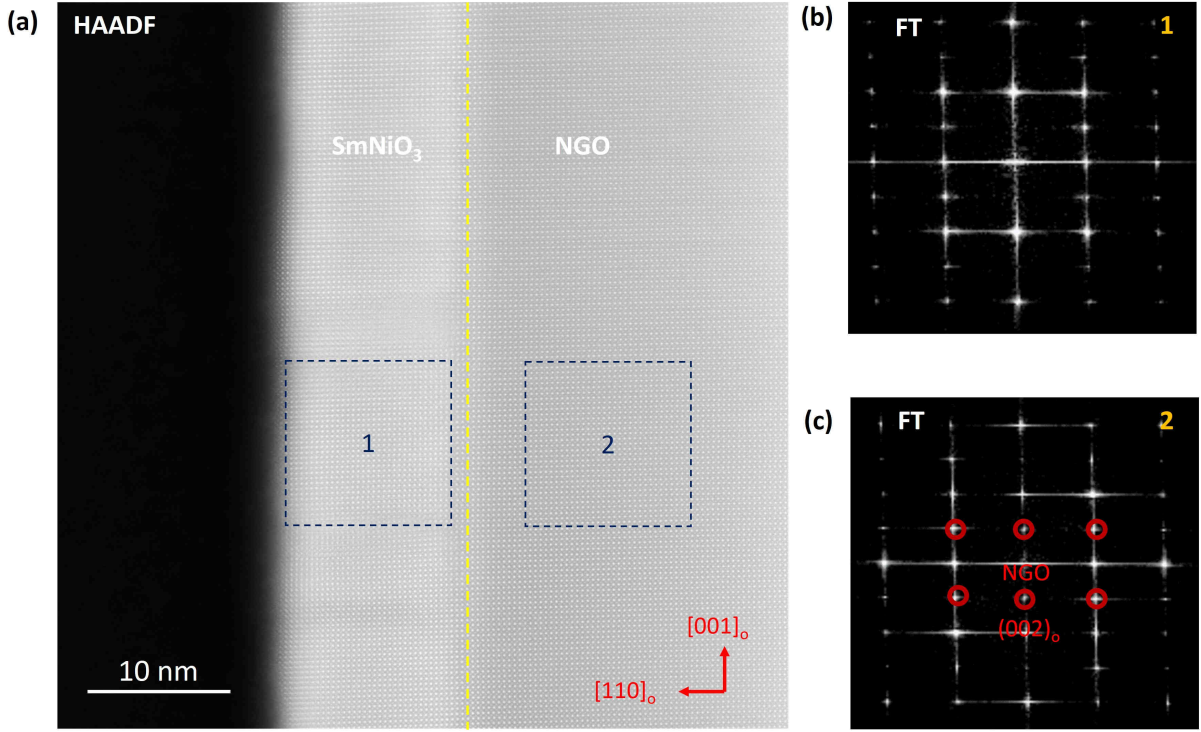

**Figure S2:** The Orthorhombic SmNiO$_3$ thin-film. (a) A low magnification HAADF image of SmNiO$_3$ thin film on the orthorhombic NdGaO$_3$ substrate with in-plane [001]$_o$ axis. (b & c) The Fourier Transform on (b) SmNiO$_3$ and (c) Substrate NdGaO$_3$. c-axis continuity, octahedral rotation, oxygen octahedral continuity.

## 2  Structural - Ab initio

The Figures S3-S6 and Table S1 consists of the ab-initio calculations that lead to the structural identification of this new phase. Beginning with Fig.S3 and S4, where a new structural family of $A_nB_nO_{3n-1}$ formed by (n0n)$_{pc}$ oxygen vacancies on a cubic perovskite ABO$_3$ is proposed. Additionally, the influence of the substrate is emphasized in Fig.S3g & h. The c-axis parameter is around 3.65Å for n = 3, and it converges to about 3.75Å for n higher than 4. Fig.S4 shows the relaxed structures of this family for n = 1 to 4. The n = 1 case corresponds to that of an infinite-layer nickelate, with c = 3.28 Å. On observing along one axis (the 0° or 90° projection mentioned in the main text), the Sm$_9$Ni$_9$O$_{22}$ have an identical structure as the A$_3$B$_3$O$_8$ in Fig. S4c, which is the n = 3 member.

The $\frac{1}{3}$ removal of oxygen can be done for a cubic system in different ways, resulting in completely different structures. This is shown in Fig. S5. The structures range from Ni being in octahedral + square-planar, octahedral + pyramidal, and octahedral + pyramidal + square-planar coordinations with the neighbouring oxygen atoms. Only the A$_9$B$_9$O$_{22}$ phase that has Ni sites in octahedral + pyramidal + square-planar coordinations as shown in Fig. S5c. Having demonstrated these for a 3x3x3 supercell of a cubic perovskite, the proper relaxation of the Sm$_9$Ni$_9$O$_{22}$ phase is carried out from a 6x6x6-supercell. This is shown in Fig. S6. Beginning with a 6x6x6-cubic supercell of SmNiO$_3$, the (303)$_{pc}$ and (033)$_{pc}$ families of apical oxygen were removed (Fig. S6a-b), reduced the c-axis parameter to 3.63 Å and the crystal was allowed to relax minimizing the total force. The resulting 960-atom tetragonal structure in different orientations is shown in Fig. S6d-f. This was then reduced to a smaller monoclinic unit cell using symmetry. The structural parametrs of this monoclinic structure is given in Table S1. It is to note that, from the symmetry, one can go back to the 960-atom tetragonal structure with the lattice vectors $\vec{a_T}$, $\vec{b_T}$, $\vec{c_T}$ exprerssed in the symmetrized monoclinic parameters ($\vec{a_m}$, $\vec{b_m}$, $\vec{c_m}$) is: $\vec{a_T} = \vec{a_m} + \vec{b_m} + 2\vec{c_m}$, $\vec{b_T} = \vec{a_m} - \vec{b_m} + 2\vec{c_m}$, $\vec{c_T} = 2\vec{a_m} - 2\vec{c_m}$.

24

S3

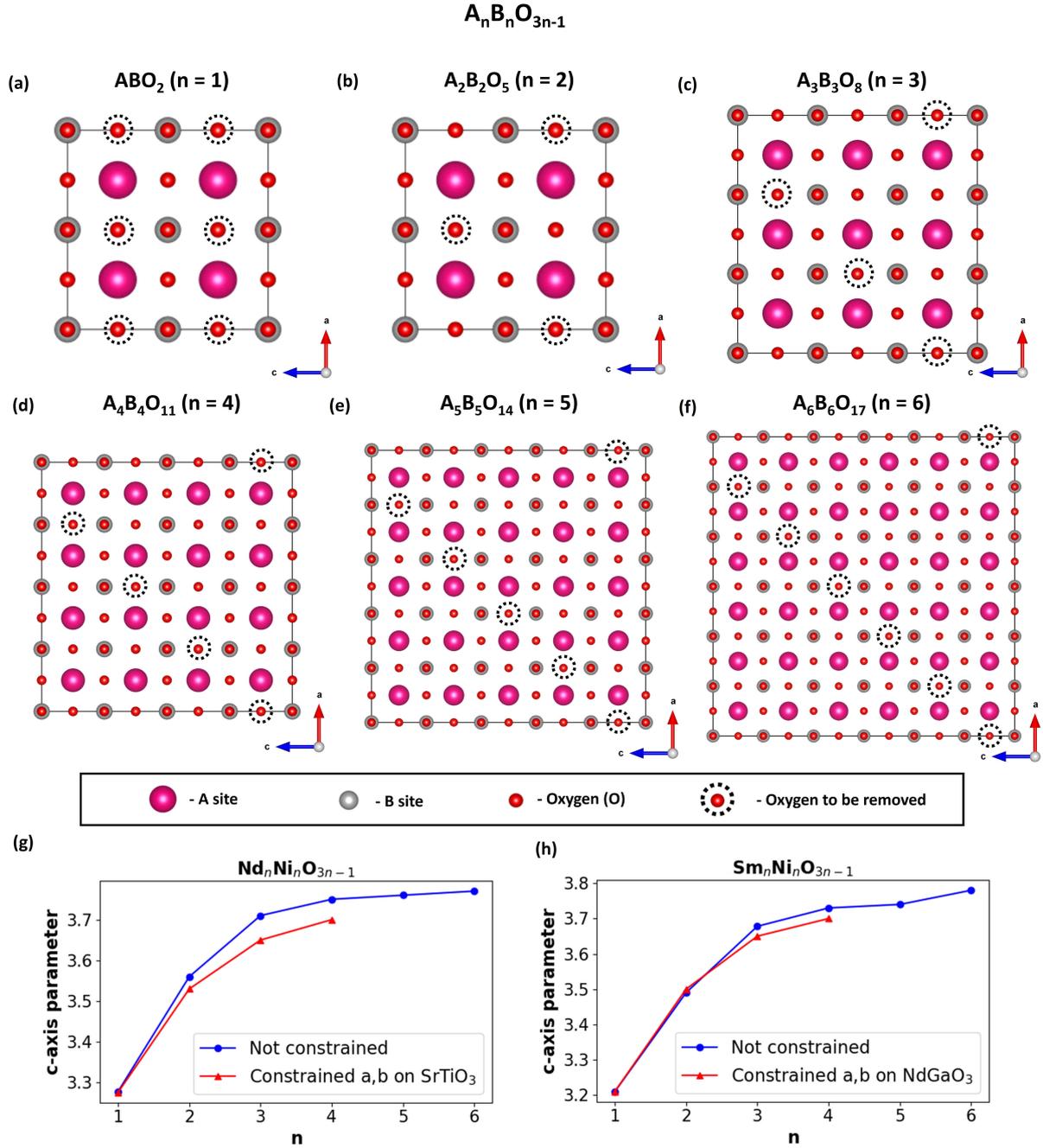

**Figure S3:** $A_nB_nO_{3n-1}$ family with the oxygen vacancies to be induced for (a-f) n = 1 - 6. (g,h) The corresponding c-axis parameters obtained by atomic relaxation for different n, with an without constraining in plane parameters on the substrate for a (g) $Sm_nNi_nO_{3n-1}$//NdGaO$_3$ system and (h) $Nd_nNi_nO_{3n-1}$//SrTiO$_3$ system. As it is evident, the c-axis parameters increases with the oxygen content ($A_nNi_nO_{3n-1}$). IL-nickelate corresponds to n = 1, while the parent perovskite would be for n = $\infty$. Furthermore constraining the in-plane parameters (NdGaO$_3$ : $a = 3.86$ Å, $b = 3.86$ Å and SrTiO$_3$ : $a = b = 3.91$ Å) with these substrates reduces the c-axis parameter upon crystal relaxation, for n higher than 2.

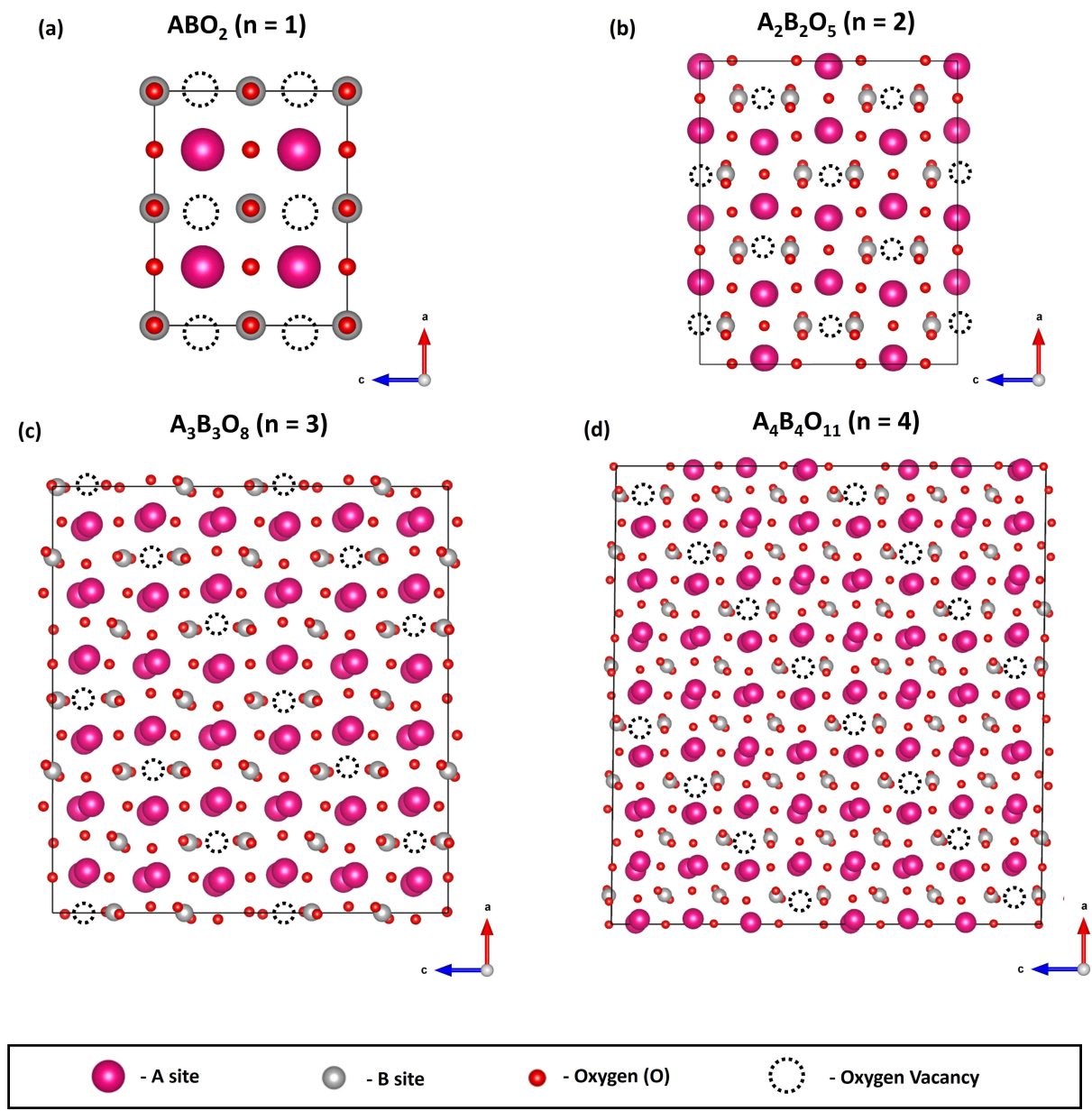

**Figure S4:** Relaxed structures of the $Sm_nNi_nO_{3n-1}$ family with the oxygen vacancies (a-d) n = 1 - 4.

26

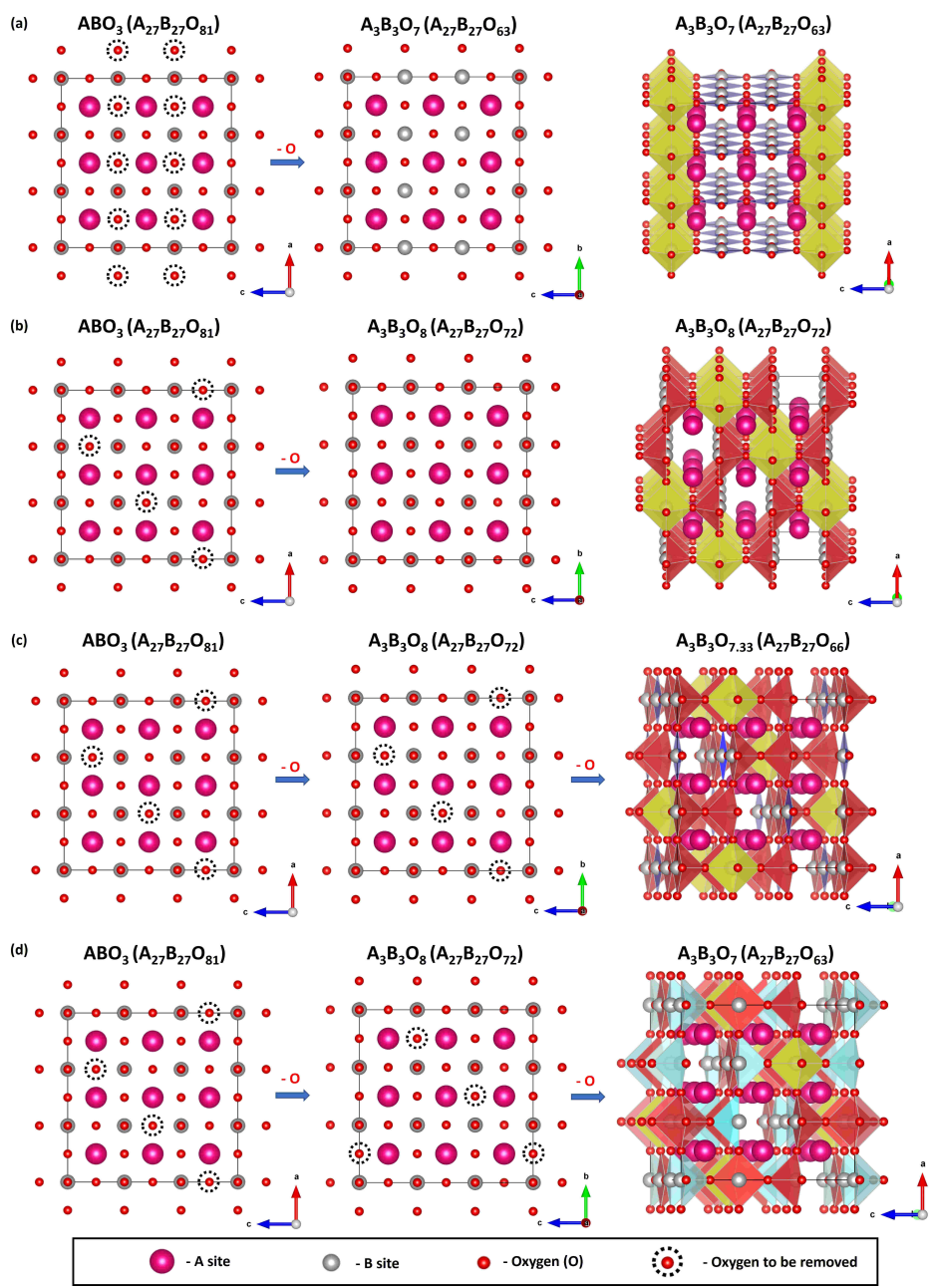

**Figure S5:** Different structures resulting by 1/3 removal of oxygen along various directions from a parent $ABO_3$ cubic 3x3x3 supercell. (a) The $A_3B_3O_7$ structure resulting from removing oxygen from the apical sites on one face as shown. It results in Ni sites existing in square-planar and octahedral coordinations with the oxygen atoms. (b) The $A_3B_3O_8$ structure resulting from removing oxygen from the apical sites with a $(303)_{pc}$ periodicity on one face. This results in Ni sites existing in pyramidal and octahedral coordinations with the oxygen atoms. (c) The $A_3B_3O_{7.33}$ ($A_9B_9O_{22}$) structure resulting from removing oxygen from the apical sites with a $(303)_{pc}$ and $(033)_{pc}$ periodicity. This results in Ni sites existing in pyramidal, square-planar and octahedral coordinations with the oxygen atoms. (d) The $A_3B_3O_7$ structure resulting from removing oxygen from the apical (out-of-plane) sites with a $(303)_{pc}$ periodicity on one face and equatorial (in-plane) sites with the same periodicity along the other face. This results in Ni sites existing in pyramidal, tetrahedral and octahedral coordinations with the oxygen atoms.

27

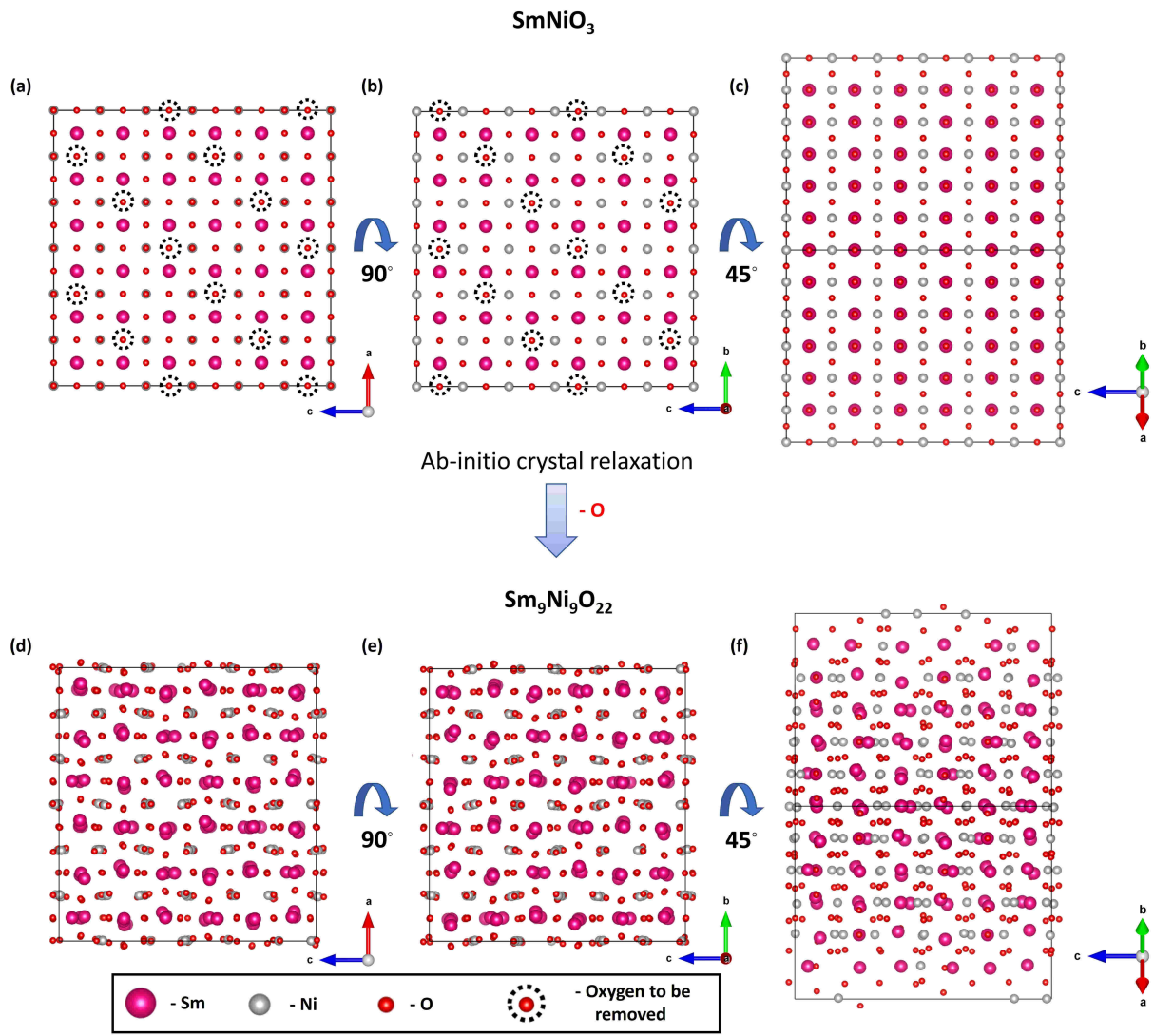

**Figure S6:** Ab-initio crystal structure relaxation of a 6x6x6 cubic supercell of SmNiO$_3$ with (303)$_{pc}$ and (033)$_{pc}$ oxygen vacancies, resulting in the SmNiO$_{2.44}$ (Sm$_9$Ni$_9$O$_{22}$) structure. (a-c) The intial 6x6x6 supercell of cubic SmNiO$_3$ ($a = 3.79$ Å), along the (a) 0°, (b) 90°, and (c) 135° projections. Note that the oxygen vacancies to be induced are indicated in (a) and (b), but they are not indicated along the 135° projection in (c) due to overlap with other oxygen atoms. (d-f) The relaxed tetragonal structure along the (a) 0°, (b) 90°, and (c) 135° projections. The in-plane parameters ($a = b = 3.79$ Å) are kept the same, while the out-of-plane parameter is reduced ($c = 3.63$ Å).

| C2/m | $a = 9.02$Å | $b = 16.09$Å | $c = 6.47$Å |
|---|---|---|---|
| (12) | $\alpha = 90.00°$ | $\beta = 87.57°$ | $\gamma = 90.00°$ |
| **Atom** | x | y | z |
| **Ni1** | 0.13629 | 0.00000 | -0.12659 |
| **Ni2** | -0.50000 | 0.00000 | 0.50000 |
| **Ni3** | 0.31121 | -0.16600 | 0.19587 |
| **Ni4** | 0.00000 | -0.16814 | 0.50000 |
| **Sm1** | 0.00000 | -0.18474 | 0.00000 |
| **Sm2** | -0.31539 | -0.15216 | 0.29485 |
| **Sm3** | 0.34038 | -0.50000 | 0.64052 |
| **Sm4** | 0.00000 | -0.50000 | 1.00000 |
| **O1** | 0.09218 | -0.08261 | -0.33878 |
| **O2** | -0.18560 | -0.08740 | -0.06043 |
| **O3** | -0.63850 | -0.08647 | 0.39467 |
| **O4** | 0.25000 | -0.25000 | 0.00000 |
| **O5** | -0.09169 | -0.24204 | 0.31874 |
| **O6** | -0.36078 | 0.00000 | 0.27300 |
| **O7** | -0.50000 | -0.14450 | 1.00000 |

**Table S1:** The structural parameters of the $Sm_9Ni_9O_{22}$ monoclinic $C2/m$ unit cell, reduced using symmetry from the relaxed 6x6x6 supercell. All the sites have an occupancy of 1.

# 3 Spectroscopy

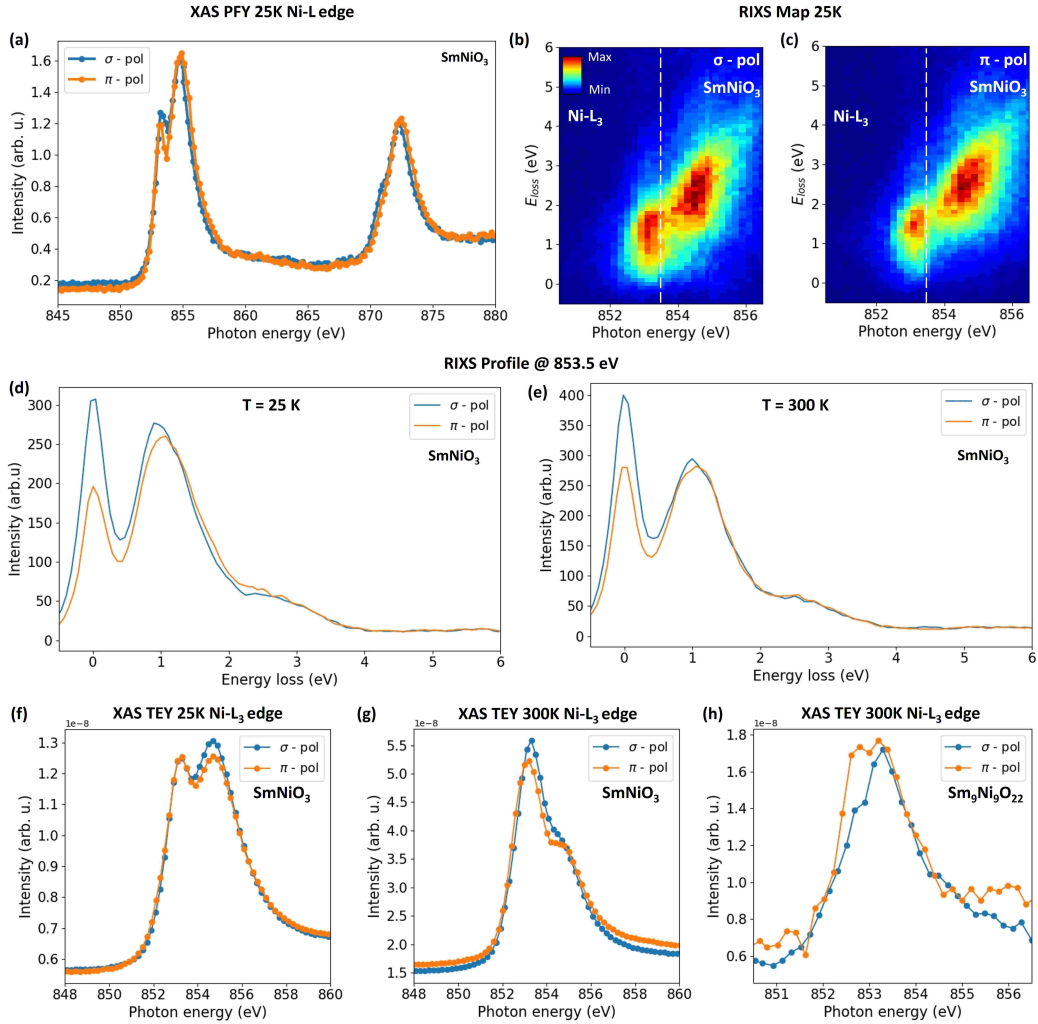

**Figure S7:** RIXS and XAS of SmNiO$_3$ and Sm$_9$Ni$_9$O$_{22}$. (a) Partial fluorescence yield (PFY) XAS spectra at Ni - L edge of SmNiO$_3$ in two incident photon polarizations at 25K. (b & c) The corresponding RIXS maps at Ni - L$_3$ of SmNiO$_3$ in (b) $\sigma$ polarization and (c) $\pi$ polarization. (d & e) Temperature dependent variations in RIXS XLD profile from both polarizations in SmNiO$_3$ at 853.5 eV for (d) T = 25 K and (e) T = 300 K. (f & g) Total electron yield (TEY) XAS spectra at Ni - L$_3$ edge of SmNiO$_3$ for two incident polarizations at (f) T = 25 K and (g) T = 300 K. (f) TEY XAS spectra at Ni - L$_3$ edge of Sm$_9$Ni$_9$O$_{22}$ for two incident polarizations at 300 K. The XAS spectra in PFY and TEY mode are identical, as depicted here for SmNiO$_3$. The TEY spectra of SmNiO$_3$ at 300K and 25K shows the metal-to-insulator transition, while the spectra of Sm$_9$Ni$_9$O$_{22}$ stays the same at 25K and 300K. Only TEY of the Ni-L are probed at 25K and 300K but they are very similar as the one reported by Bisogni et al. [1] showing that part of the thin film, certainly close to the surface is entering in the MIT coexistence. At RT, indeed the peak separation is getting smaller, and the width is changing very similarly as what they report by TEY in the case of NdNiO3. RIXS probes more bulk and hence shows weaker differences. Nevertheless, a small shift to higher energy is seen at $\approx$ 1 eV energy loss measured at an excitation energy of 853.5 eV. It shows, at least, a partial charge disproportion when compared to the RIXS reported by Bisogni et al. [1].

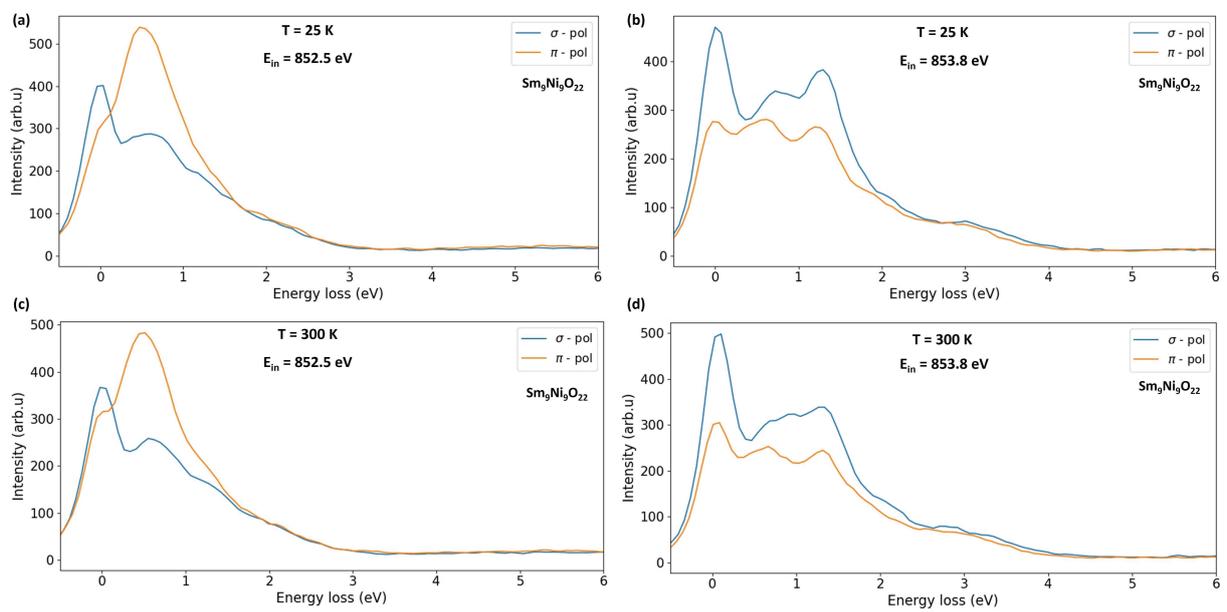

**Figure S8:** Temperature dependant RIXS line Profile of $Sm_9Ni_9O_{22}$ at different photon energies. (a & b) The measurements at 25K, for a photon energy of (a) 852.5 eV and (b) 853.8 eV. (c & d) The measurements at 300K, for a photon energy of (c) 852.5 eV and (d) 853.8 eV.

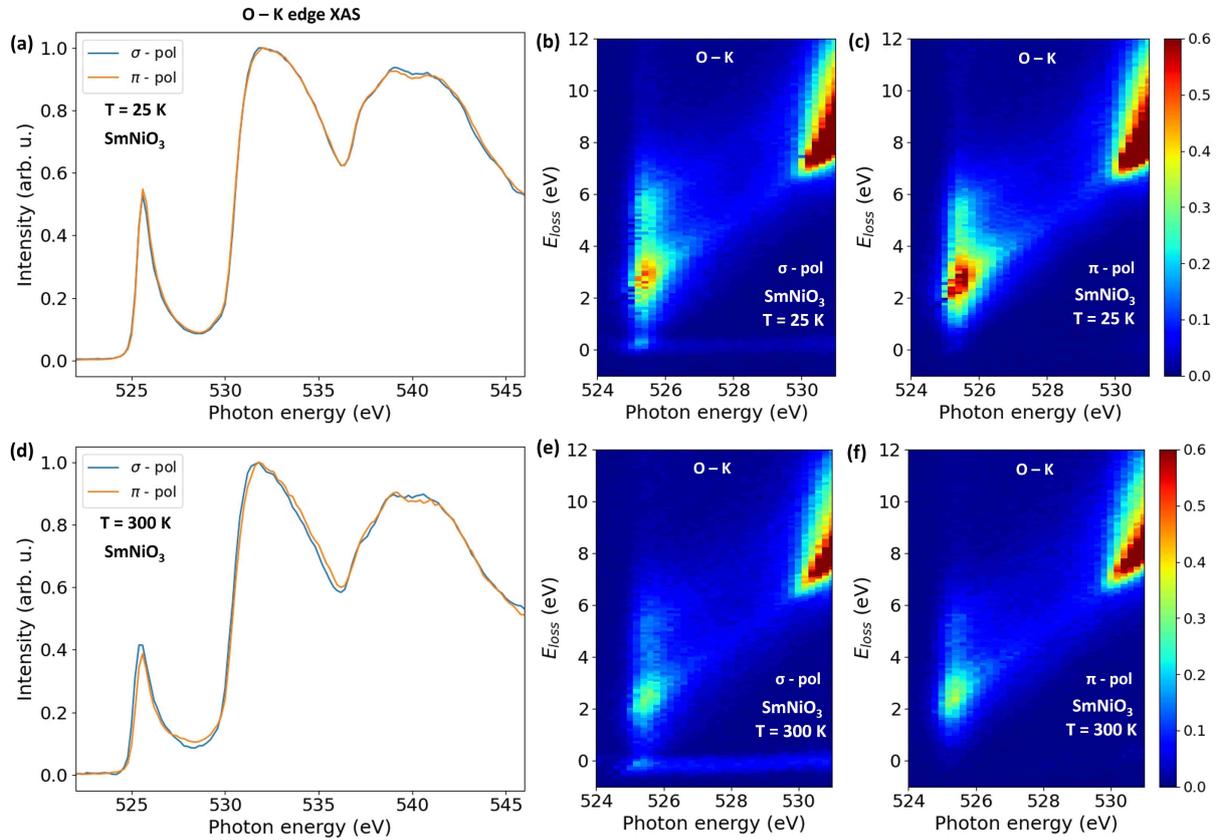

**Figure S9:** Temperature dependent spectroscopic studies at the O-K edge by XAS and RIXS (in energy loss scale) of SmNiO$_3$. (a-c) Measurements at T = 25 K, (a) O-K edge PFY XAS spectra in $\sigma$ and $\pi$ incident polarizations, (b & c) RIXS map at O-K edge for (b) $\sigma$ incident photon polarization and (c) $\pi$ incident photon polarization. (d-f) Measurements at T = 300 K, (d) O-K edge PFY XAS spectra in $\sigma$ and $\pi$ incident polarizations, (e & f) RIXS map at O-K edge for (e) $\sigma$ incident photon polarization and (f) $\pi$ incident photon polarization. Notably, as seen in XAS and made even clearer in the RIXS maps, the energy width of the O-K pre-edge is getting narrower and is shifted to a higher absorption energy at low temperature. Similar narrowing at the XAS pre-edges was reported for a nickelate metal-to-insulator transition (MIT) [2], confirming that some domains have a metallic character at RT. As expected, the parent structure has a weak dichroism for its spectral signatures, including the Raman-like energy losses observed at an excitation energy corresponding to the O-K pre-edge. In this pre-edge region, RIXS maps also evidence low-energy loss features (less than 0.5 eV) only at RT, in a rather similar way that was reported with low-energy electron-hole pair continuum at the Ni-L edge RIXS for a metallic nickelate [1].

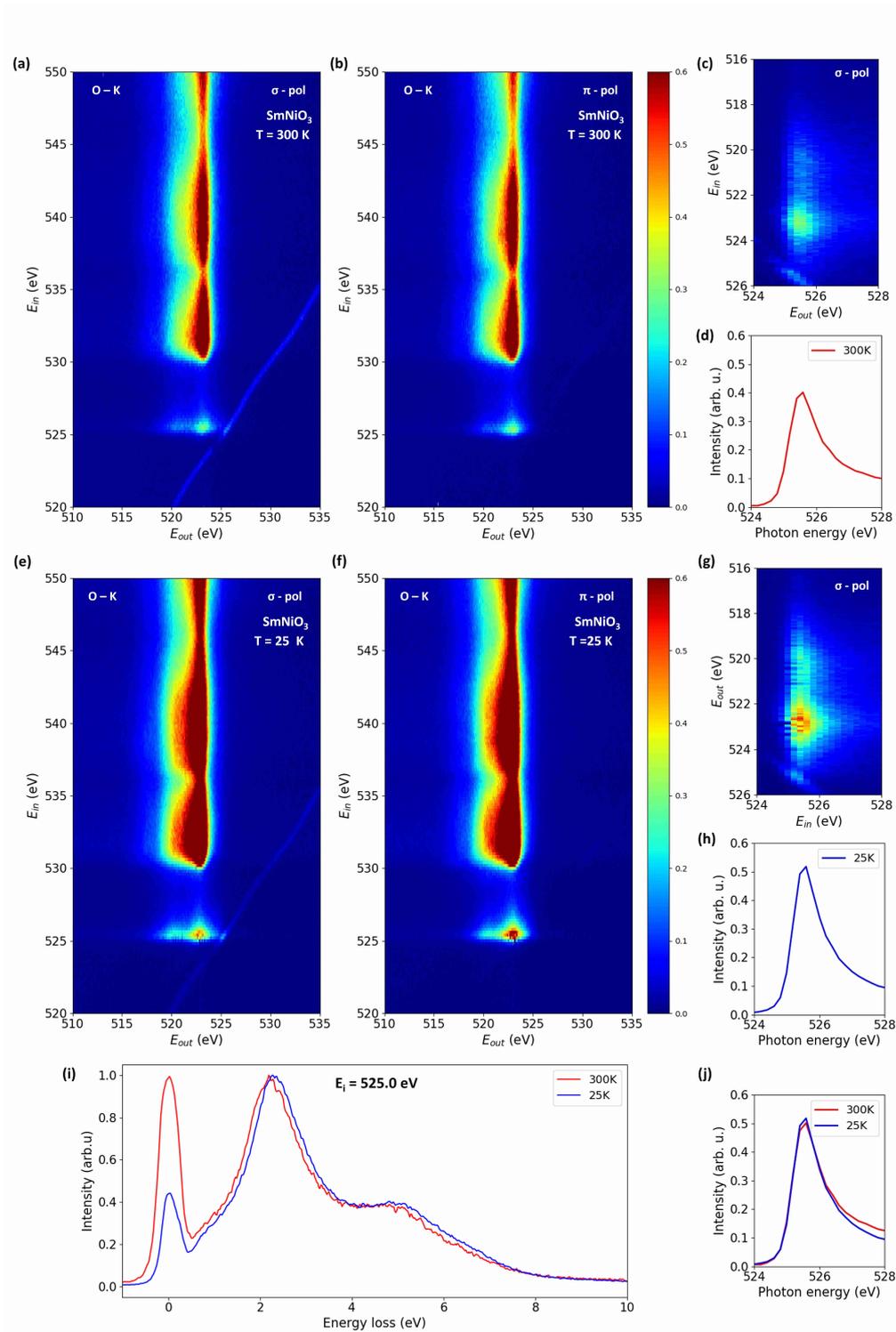

**Figure S10:** RIXS Map and XAS of O-K edge of SmNiO$_3$ showing a temperature dependent electronic transition. (a-d) Measurements at 300K. (a,b) RIXS map in incident (E$_{in}$) and detected (E$_{out}$) photon energy scales for (a) $\sigma$ and (b) $\pi$ incident photon polarisations, (c) A magnified RIXS map of a selected region in (a), (d) The XAS PFY plot of the O-K pre-edge, combining both polarisations. (e-h) Measurements at 25K. (e,f) RIXS map in incident (E$_{in}$) and detected (E$_{out}$) photon energy scales for (e) $\sigma$ and (f) $\pi$ incident photon polarisations, (g) A magnified RIXS map of a selected region in (e), (h) The XAS PFY plot of the O-K pre-edge, combining both polarisations. (i) Comparing the RIXS line profile at an incident photon energy of 525.0 eV for temperatures 300K and 25K, showing a blue shift at 300K. (j) The normalized-maxima-aligned pre-edge of O-K, for both temperatures. It shows a wider pre-edge at 300K.

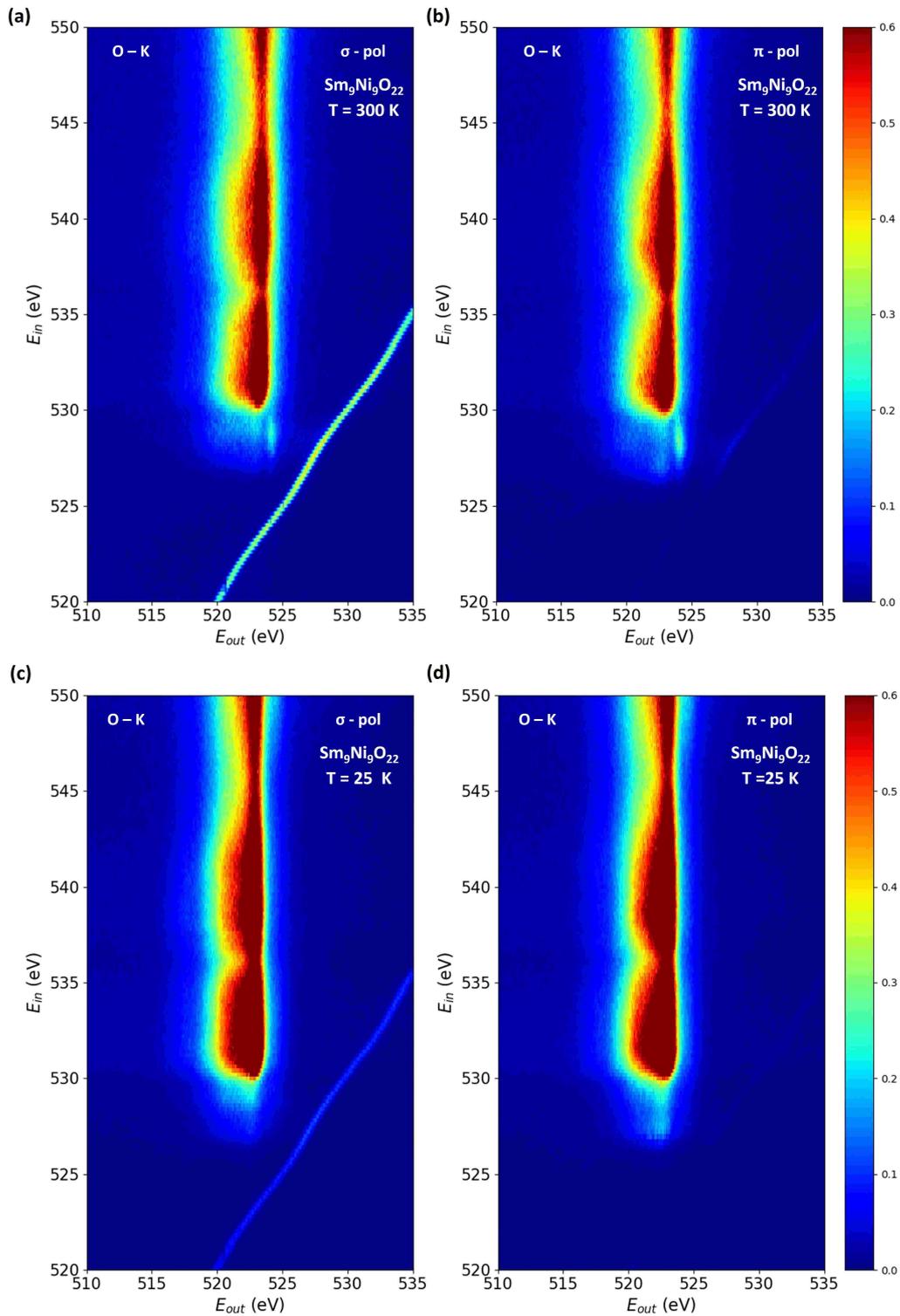

**Figure S11:** RIXS Map of O-K edge of $Sm_9Ni_9O_{22}$ showing the electronic transition at different temperatures. (a,b) RIXS map in incident ($E_{in}$) and detected ($E_{out}$) photon energy scales for (a) $\sigma$ and (b) $\pi$ incident photon polarisations at 300K, (c,d) RIXS map in incident ($E_{in}$) and detected ($E_{out}$) photon energy scales for (c) $\sigma$ and (d) $\pi$ incident photon polarisations at 25K.

34

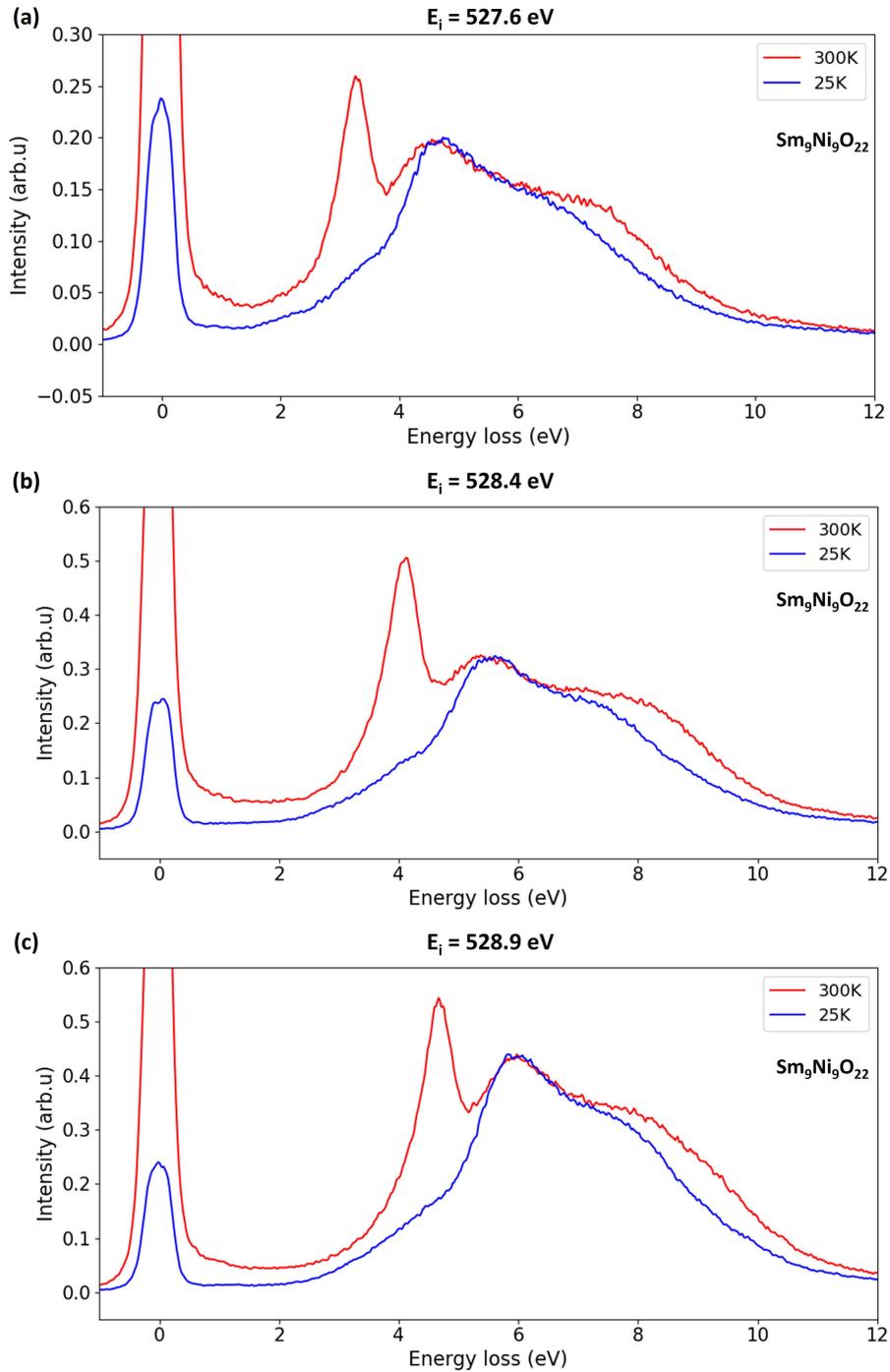

**Figure S12:** Temperature dependant RIXS line Profile of $Sm_9Ni_9O_{22}$ at different photon energies for the $\sigma$ polarised incident photons. The spectra at an incident photon energy of (a) 527.6 eV, (b) 528.4 eV, and (c) 528.9 eV.

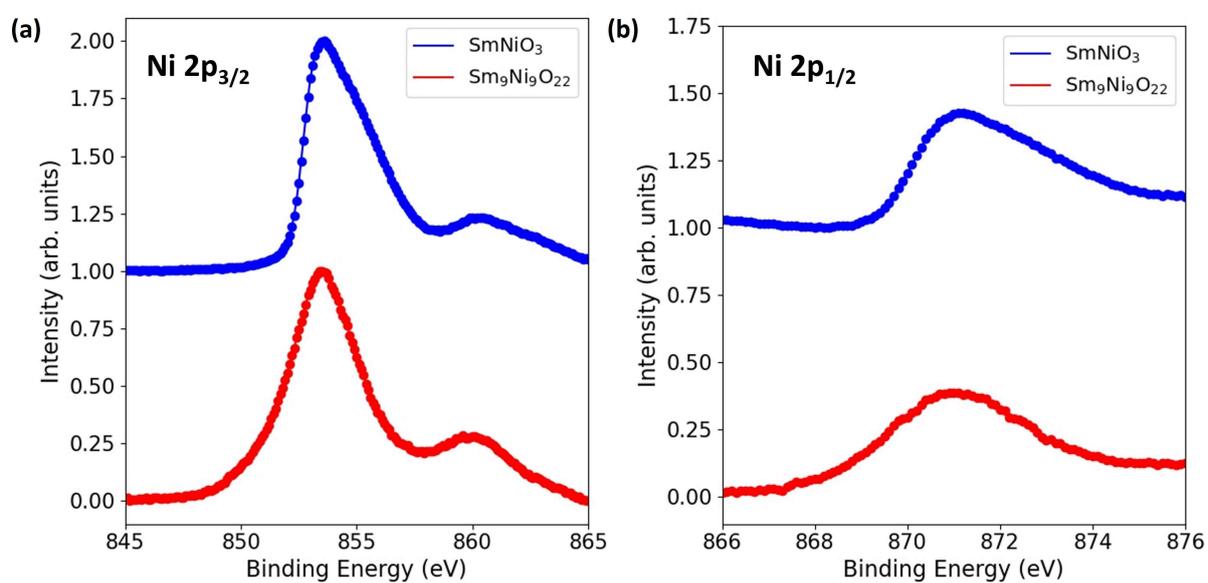

**Figure S13:** HAXPES comparison of $Sm_9Ni_9O_{22}$ and $SmNiO_3$ for (a) Ni $2p_{3/2}$ and (b) Ni $2p_{1/2}$ core levels. The measurements were done in a bulk sensitive mode, probing the whole thin-film. In both, $Sm_9Ni_9O_{22}$ has a low binding energy (LBE) shoulder, indicative of the $Ni^{1+}$ in the square-planar sites. The bulk-probing capacity of HAXPES makes it insensitive to possible surface contaminations or artefacts. One noticeable difference is the presence of a low binding energy (LBE) shoulder for $Sm_9Ni_9O_{22}$ compared to $SmNiO_3$ that can also be attributed to the $Ni^{2+}$ and $Ni^{1+}$ sites.

# 4 Electronic - Calculations

The orbital occupancy, net spin polarization, and density of states of $Sm_9Ni_9O_{22}$ were calculated using DFT+U framework of Quantum ESPRESSO code [3,4]. The DFT+U calculations were performed with a J = 1.2 eV, making $U_{effective}$ = U - J, on the symmetrized monoclinic $Sm_9Ni_9O_{22}$ unit cell given in Table S1. The Ni 3d orbital occupancy obtained for a U of 6 eV and a J of 1.2 eV is given in Table S2. It evidences the square-planar Ni site to be in a $3d^9$, pyramidal and octahedral Ni sites to be in a $3d^8$ occupancies. The corresponding spin-polarized density of states is shown in Fig. S16. The multiplet calculations were performed using CTM4XAS, CTM4RIXS [5], and Quanty [6]. The calculation for a D4h $Ni^{2+}$ was done in CTM4XAS and CTM4RIXS using the crystal field parameters $D_q = 0.13$ eV, $\Delta_t = 0.07$ eV, $\Delta_s = 0.15$ eV. The dichroic XAS results for the Ni-L edge are shown in Fig. S14, and the RIXS results are shown in Fig.4 in the main text. Additionally, the RIXS map for a D4h $Ni^{1+}$ was done using Quanty, with $D_q = 0.13$ eV, $\Delta_t = 0.10$ eV, $\Delta_s = 0.30$ eV for the crystal-field parameters. The results are shown in Fig.S15.

| Ni site | Spin | $d_{x^2-y^2}$ | $d_{z^2}$ | $d_{xy}$ | $d_{xz}$ | $d_{yz}$ |
|---|---|---|---|---|---|---|
| Square-planar | up | 0.81 | 0.98 | 0.91 | 0.98 | 0.98 |
| | down | 0.16 | 0.97 | 0.77 | 0.97 | 0.98 |
| Pyramidal | up | 0.96 | 0.97 | 0.99 | 0.99 | 0.99 |
| | down | 0.17 | 0.20 | 0.98 | 0.98 | 0.99 |
| Octahedral | up | 0.98 | 0.98 | 0.99 | 0.99 | 0.99 |
| | down | 0.13 | 0.16 | 0.98 | 0.98 | 0.98 |

**Table S2:** Occupation of 3d orbital in different Ni sites in $Sm_9Ni_9O_{22}$ by DFT calculations with a Hubbard U parameter of 6 eV and Hund's J parameter of 1.2 eV. The absolute values of the results indicates a $3d^8$ occupancy for the pyramidal and octahedral Ni sites, with a $3d^9$ occupancy for the square-planar sites.

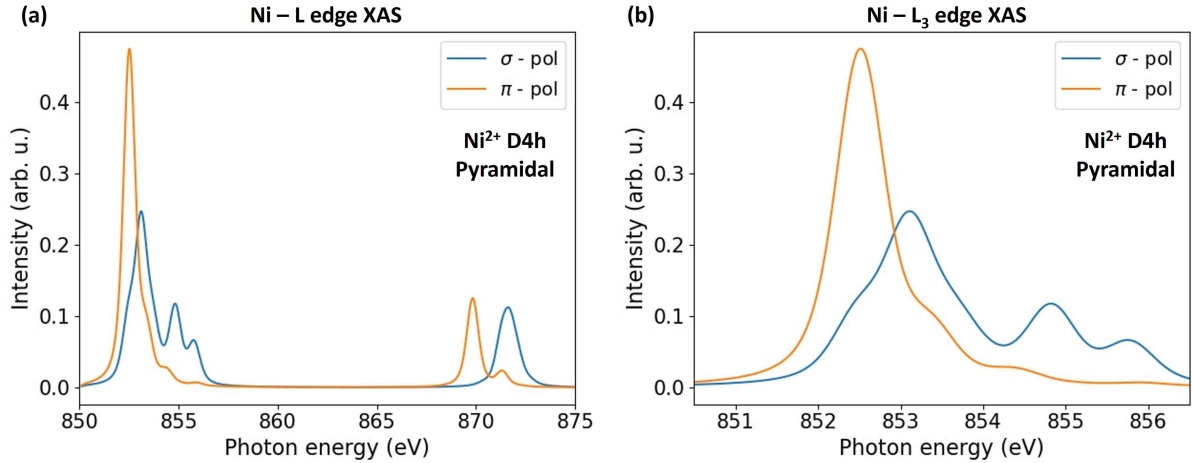

**Figure S14:** CTM4XAS [5] calculation of $Ni^{2+}$ in D4h for the pyramidal sites. (a) Results for both polarizations for the Ni-L edge, (b) a magnified Ni-$L_3$ edge for both polarizations. It has been obtained using $D_q = 0.13$ eV, $\Delta_t = 0.07$ eV, $\Delta_s = 0.15$ eV for the crystal-field parameters. Same parameters are used for the RIXS calculations. It gives the following orbital energetic hierarchy : $E_{x^2-y^2}$(b1g) = 0 eV (reference) ; $E_{z^2-r^2}$(a1g) = - 0.95 eV ; $E_{xz/yz}$(eg)= - 1.30 eV ; $E_{xy}$(b2g)= - 1.40 eV

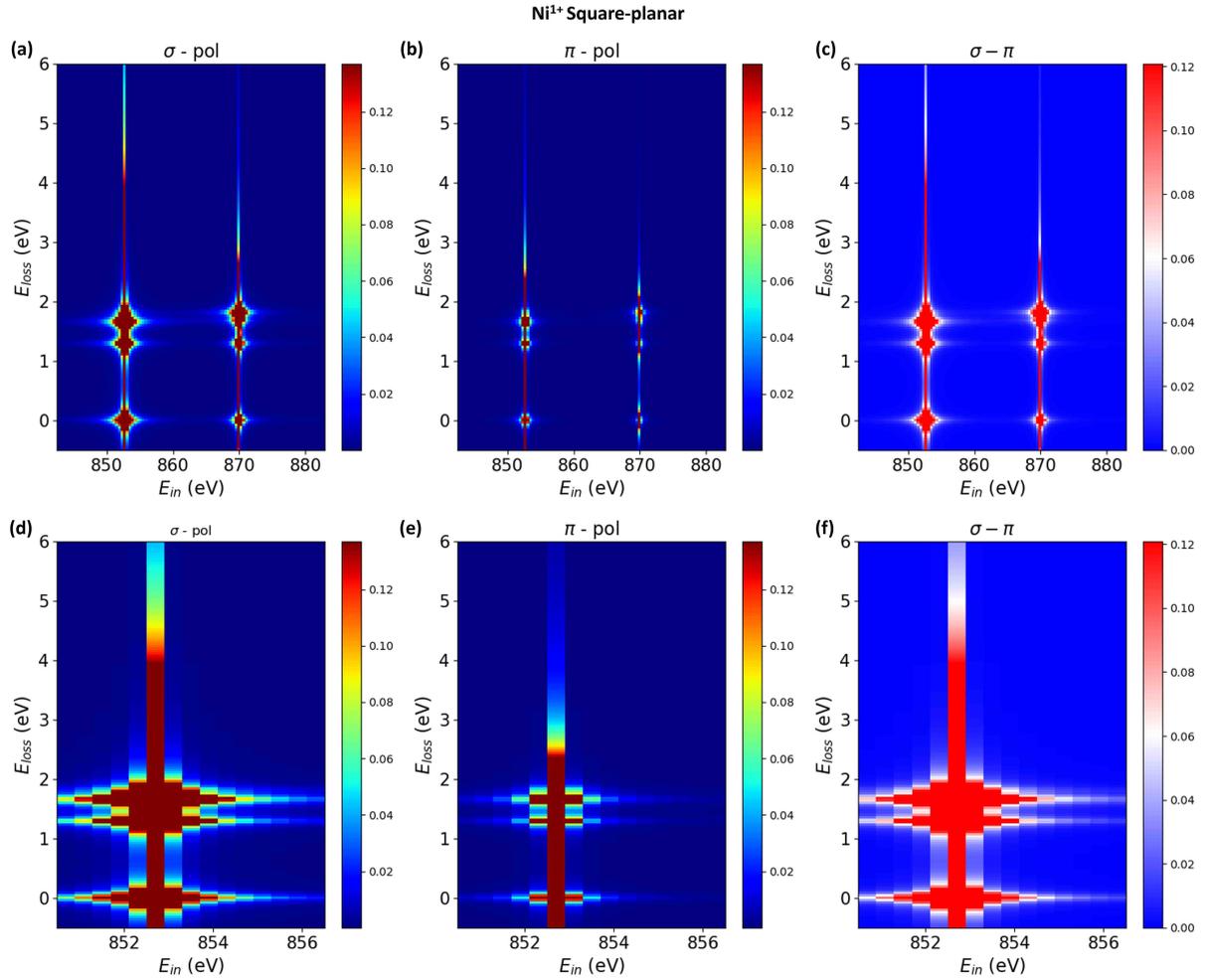

**Figure S15:** Calculated RIXS map at Ni-L edge of Ni$^{1+}$ Square-planar sites by Quanty [6]. (a) Results for $\sigma$ polarization, (b) $\pi$ polarization, (c) Difference of both polarization ($\sigma - \pi$). (d-f) The magnified map at the Ni-L$_3$ edge. It has been obtained using $D_q = 0.13$ eV, $\Delta_t = 0.10$ eV, $\Delta_s = 0.30$ eV for the crystal-field parameters. It gives the following orbital energetic hierarchy : E$_{x^2-y^2}$(b1g) = 0 eV (reference) ; E$_{xy}$(b2g)= - 1.30 eV ; E$_{z^2-r^2}$(a1g) = - 1.70 eV ; E$_{xz/yz}$(eg)= - 1.70 eV.

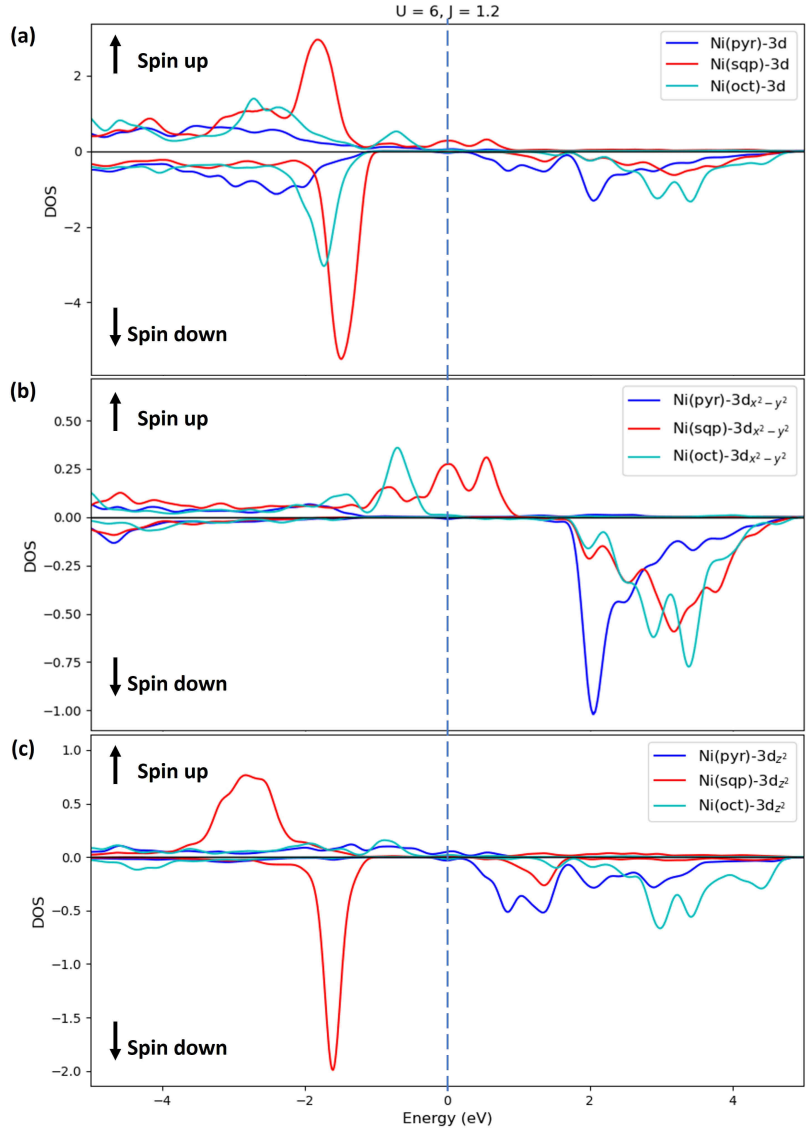

**Figure S16:** Partial density of states of different Ni sites in $Sm_9Ni_9O_{22}$ from DFT+U calculations, for U = 6 and J = 1.2 eV for (a) Total 3d orbital, (b) $3d_{x^2-y^2}$ orbital, and (c) $3d_{z^2}$ orbital. A first approximation is to consider the O-K edge as a probe of the unoccupied densities of state with O-2p character. The O-K pre-peak is then arising from the hybridization of the O-2p with the Ni-d band. When comparing the partial density of states above the Fermi level, alternating sequence of in-plane ($3d_{x^2-y^2}$), out-of-plane ($3d_{z^2}$) and in-plane ($3d_{x^2-y^2}$) states are legible, in agreement with the $\sigma$ and $\pi$ dependence of the XAS measurement.


\end{document}